\documentstyle[aps,eqsecnum,epsf]{revtex}

\begin{document}
\thispagestyle{empty}

\begin{flushright}
JLAB-THY-97-07 \\
February 27, 1997 \\ 
hep-ph/9702443
\end{flushright}

\begin{center}
{\Large \bf Transverse  Momentum and Sudakov 
Effects \\
in Exclusive QCD Processes:  
 $\gamma^* \gamma\pi^0$ Form Factor}
\end{center}
\begin{center}
{I.V. MUSATOV,  A.V. RADYUSHKIN
\footnotemark 
}  \\
{\em Physics Department, Old Dominion University,}
\\{\em Norfolk, VA 23529, USA}
 \\ {\em and} \\
{\em Jefferson Lab, 
  Newport News, VA 23606, USA}
\end{center}
\vspace{2cm}

\footnotetext{Also Laboratory of Theoretical Physics, 
JINR, Dubna, Russian Federation}

\begin{abstract}

We analyze effects due to transverse degrees of freedom in QCD calculations
of the fundamental hard exclusive amplitude of $\gamma^*\gamma \to  \pi^0$ 
transition. A detailed  discussion is given of the relation between the  
modified factorization approach (MFA)  of Sterman {\it et al.}  and standard 
factorization (SFA). Working in Feynman gauge, we  construct basic building 
blocks of MFA from the one-loop coefficient function of the SFA, 
demonstrating that Sudakov effects are distinctly different from higher-twist 
corrections. We show also that the handbag-type diagram, contrary to naive  
expectations, does not contain  an  infinite chain of $(M^2/Q^2)^n$ 
corrections: they come only from  diagrams with  transverse gluons 
emitted from the hard propagator. A simpler picture emerges within the 
QCD sum rule approach: the sum over soft  $\bar q G \ldots G q$ Fock 
components is  dual to  $\bar qq$ states generated by the local axial 
current.  We combine the  results based on QCD sum rules with pQCD 
radiative corrections and observe that the gap between our  curves  for 
the asymptotic and CZ distribution amplitudes is sufficiently large 
for an  experimental discrimination between them.

\end{abstract}

\newpage 

\section{Introduction}

The form factor $F_{\gamma^* \gamma^*  \pi^0}(q_1^2,q_2^2)$  relating two  
(in general, virtual)   photons with the lightest hadron, the pion, 
plays a crucial role in  the studies 
of  exclusive processes in quantum chromodynamics. 
With  only one hadron involved, it has 
the simplest structure analogous
to that of the  form factors
of deep inelastic scattering.
At large photon virtualities,
comparing the pQCD predictions \cite{bl80,bhl,blin,AuCh81,braaten,kmr} 
with experimental data,
one can get important information about the shape 
of the pion distribution amplitude $\varphi_{\pi}(x)$.
Due to its relation to axial anomaly
\cite{anomaly},  the $\gamma^* \gamma^*  \pi^0$ form factor
was an object of intensive studies 
since the 60's \cite{cornwall,gt,bkt,pk,ter,jacowu}. 
Experimentally, 
 $F_{\gamma^* \gamma^*  \pi^0}(q_1^2, q_2^2)$ 
for small virtuality of one of the photons,
$q_1^2\approx 0$,  was measured only recently at 
$e^+e^-$ colliders by CELLO \cite{cello}
and CLEO \cite{cleo} collaborations
(in the latter case, only a preliminary
announcement of the results was made).
The possibility  to measure  
$F_{\gamma^* \gamma^*  \pi^0}(q_1^2 \approx 0, q_2^2)$ 
at fixed-target machines like CEBAF of  Jefferson Lab
was also discussed \cite{afanas}.
These measurements  inspired the studies of the 
 momentum dependence of  this form factor 
within  various models
of the nonperturbative quark dynamics
\cite{ggmikh,hiroshi,misha,frank,kroll,pl,apa95,rr,ajohlule,huang,anis}.

For a detailed comparison of pQCD 
predictions with experimental data, one should 
have  reliable estimates of possible corrections to the 
lowest-order handbag contribution, in particular, those due to the
gluon radiation and 
 higher twist effects.
Within the standard pQCD factorization approach,
the one-loop radiative corrections to the coefficient function
were calculated in refs.\cite{AuCh81,braaten,kmr}.
  The authors of refs. \cite{kroll,ong} incorporated
the modified factorization  approach
 of Sterman and collaborators \cite{bottssterman,listerman} in  which the 
factorization  formula invloves an extra integration
over the impact parameter $b_{\perp}$ 
and  Sudakov  double logarithms of  $(\alpha_s\ln^2(b_{\perp}^2))^n$ type
are  summed to all orders. 
In refs. \cite{kroll,ong} it was  claimed that 
such an analysis takes into account
some  transverse-momentum effects neglected within
the standard factorization approach \cite{bl80,tmf78,pl80,czpr}. 
Incorporating  the transverse-momentum-dependent 
wave function $\Psi(x,k_{\perp})$,
Jakob {\it et al.} \cite{kroll} also  proposed a 
model for the effects due to the intrinsic 
(primordial) transverse momentum.

Another attempt to take into account 
the transverse momentum effects was made
by Cao  {\it et al.} \cite{huang} where the light-cone 
formalism expression \cite{bl80} 
for the $\gamma^*\gamma \to \pi^0$
 was used.  Adopting an  exponential ansatz for
 the transverse momentum dependence of the
 wave function, the authors observed 
 large ``higher-twist'' corrections,
 with the conclusion that it is difficult
 in such a situation to make a clear distinction
 between different shapes of the pion 
 distribution amplitude.

In this paper,  
 we will discuss various types of
 transverse momentum effects for the   
 $\gamma^*\gamma  \pi^0$  form factor.
 First, we briefly outline the derivation
 of the leading-twist pQCD formula for this process
 using a covariant OPE-like factorization approach \cite{20,tmf78,20a}.
 In this framework,  we identify the basic types of the higher twist
 corrections neglected in the leading-twist approximation.
 We show, in particular, that for massless quarks
in a scalar theory   no intrinsic transverse momentum
 effects are neglected in the handbag diagram:
 due to  the simple singularity structure of the
 massless quark propagator, such effects can be taken
 into account exactly and lead to negligible pion 
 mass corrections $m_{\pi}^2/Q^2$ only.
In QCD, the handbag diagram contains a  twist-4 
term interpretable as a $O(k_{\perp}^2)$ correction,
but no terms corresponding to higher  powers  of  $k_{\perp}^2$.
Hence,  the infinite tower of $(M^2/Q^2)^n$
corrections is generated by operators corresponding to higher  
$\bar q G \ldots G q$ Fock components.
In Section II, we also discuss the structure of 
the results for the one-loop radiative corrections 
\cite{AuCh81,braaten,kmr}  calculated within the standard 
factorization approach \cite{tmf78,bl80,czpr,ditrad,field}.

In section III, we give  a detailed  one-loop derivation 
of  the basic formulae of the 
modified factorization approach (MFA). 
We write the relevant one-loop integrals 
in Sudakov variables
used in \cite{bottssterman,listerman}, 
introduce the impact 
 parameter $b_{\perp}$ as the Fourier conjugate variable
 to the transverse momentum $k_{\perp}$
and  reproduce  (at one loop)  
 the structure of the modified factorization  \cite{bottssterman}.
In contrast (and complementary) to the original analysis, 
 we use  Feynman gauge 
which allows us to make a direct graph by graph
comparison with the results \cite{AuCh81,braaten,kmr} 
obtained within  the standard factorization approach (SFA). 
Since the modified factorization formulas 
appear as an intermediate step in  our 
calculations which eventually produce the results of 
the SFA,  
the two types of  factorization  give
identical results at any finite order of perturbation theory. 
 The difference between the two approaches
is  only in different organization 
of all-order summation of higher-loop terms.  
Namely, in the MFA, the Sudakov-type double logarithms
$(\alpha_s \ln^2 (Qb_{\perp} ))^n$  are treated as  logarithmic 
enhancements and are summed over  all orders 
to produce a  factor suppressing the contributions from the 
 large-$b$ region. In the standard approach, 
the $(\alpha_s \ln^2 (Qb))^n$ terms are integrated over $b_{\perp}$ 
and included order by order.  We show that for the 
$\gamma^*\gamma  \pi^0$  form factor 
the use of the SFA procedure is well justified 
since the results of the $b_{\perp}$-integration 
produce rather mild corrections ( $\sim 20\%$ at one loop).
Another lesson from our detailed one-loop study of the MFA 
is that though the factorization formula of the MFA
 explicitly  involves an  integral over the impact parameter $b_{\perp}$
(or transverse momentum $k_{\perp}$), the results 
of such an integration do not produce power suppressed 
contributions.  Thus, despite   the claims made, $e.g.,$ 
in refs.\cite{jakroll,jipaszcz,kroll}  
 higher-twist 
corrections  are not  included  in the MFA .

 In section IV, we discuss two recent attempts \cite{kroll,huang} to 
model   the intrinsic 
 momentum corrections  for the  $F_{\gamma^*\gamma  \pi^0}(Q^2)$ 
form factor. The approach of 
Jakob {\it et al.}  \cite{kroll} is based on the 
extrapolation of the modified factorization
formula into the nonperturbative region.
At large impact parameters $b$, 
the Sudakov suppression factor is supplemented
by   the nonperturbative wave function $\widetilde \Psi (x,b)$
reflecting the effects due to 
the primordial transverse momentum distribution.
However, since  terms which were inessential
for the derivation of the  Sudakov  factor  at large $Q^2$ 
may be quite important for small $Q^2$,
it is not  clear  for which $Q^2$-region
such an extrapolation is sufficiently accurate.
We observe, in particular, that instead of producing the $Q^2=0$ 
value  dictated by the axial anomaly \cite{anomaly,f0},
the extrapolation formula gives a logarithmically divergent result
suggesting that the extrapolation should not go down to very low $Q^2$.
 Cao {\it et al.} \cite{huang}
use  the expression for the
$\bar q q$ Fock state contribution to  $F_{\gamma^*\gamma  \pi^0}(Q^2)$
derived in the light-cone formalism by Brodsky and Lepage \cite{bl80}.
This expression  involves no approximations and  has correct limits  
 both for   small and  large $Q^2$. 
In particular, we demonstrate that,
in full accordance with our general analysis,
it contains no higher-twist contributions.
Still, one should take into account that
 the $\bar qq $ term, by definition, 
 does not include the contribution
due to  higher $\bar q G \ldots G q$  Fock 
components of the pion light cone wave function. 
As shown in ref.\cite{bhl}, the latter 
 coincides in the real photon limit $Q^2=0$
 with that of the $\bar qq$ Fock component
and doubles the total result at this point. 
Clearly, the  inclusion  (or at least modelling) of this contribution
is necessary for a consistent description of subasymptotic effects.
Comparing the approaches of refs.\cite{kroll,huang}, we
  emphasize that they  incorporate  two 
completely different light-cone schemes.  
The light-cone  formalism of Brodsky  and Lepage \cite{bl80}
 used in 
ref.\cite{huang}  is equivalent to incorporating 
  the  infinite momentum frame.
On the other hand, the approach of   ref.\cite{kroll}  
(and that of the underlying papers \cite{bottssterman,listerman}) 
is based on the  Sudakov decomposition. 
 The basic difference  between the two light-cone approaches 
 is that the momentum of the virtual photon in the
 $\gamma^* \gamma \to \pi^0$ process is dominated by 
the  transverse
component  in the BL light-cone  scheme while it is
  purely  longitudinal in the Sudakov approach. 
  
  In Section V,  we  use     QCD sum rule ideas
to get a model for the $F_{\gamma^* \gamma \pi^0}(Q^2)$ 
form factor  which reproduces  both the $Q^2=0$ constraint 
imposed by the axial anomaly   and  the lowest-order 
 pQCD results 
for  high $Q^2$.   We show also that 
the results obtained on the basis of QCD sum rules
and quark-hadron duality can be interpreted
in terms of the effective valence wave function
which absorbs  information about  soft 
dynamics of higher Fock components
of the standard light-cone approach.  
Combining these results with pQCD radiative corrections,
we  obtain an  expression depending on  the 
choice of the low-energy distribution amplitude.
The difference between 
our results for the asymptotic and
CZ distribution amplitudes is sufficiently large 
for an unambiguous experimental discrimination
between these two  possibilities.

\section{Factorization}

\subsection{Structure of factorization}

We define the form factor 
$F_{\gamma^*\gamma^*  \pi^o} \left(q_1^2,q_2^2\right)$
of the $\gamma^*\gamma^* \to \pi^o$ transition
through  the matrix element
\begin{equation}
4 \pi \int
\langle {\pi},\stackrel{\rightarrow}{p}
|T\left\{J_{\mu}(X)\,J_{\nu}(0)\right\}| 0 \rangle e^{-iq_1 X } d^4 X
 = i e^2 \sqrt{2}  \,\epsilon_{\mu \nu \alpha  \beta}
q_1^{\alpha} q_2^{\beta}
 F_{\gamma^*\gamma^*  \pi^0}\left(q_1^2,q_2^2 \right)
\label{eq:form}
\end{equation}
where  $J_{\mu}$\ is the electromagnetic current of the light quarks
\begin{equation}
J_{\mu}=e_u \bar{u}\gamma_{\mu}u +
e_d \bar{d}\gamma_{\mu}d  
\label{eq:tokem}
\end{equation}
and  $|  {\pi},\stackrel{\rightarrow}{p} \rangle$  is a one-pion state
with  the 4-momentum $p$. 
Note, that our definition 
(aimed at getting a simple coefficient for the spectral density
for the triangle anomaly diagram, see Section V)
differs from that  in refs.\cite{bl80,kroll,huang}
by factor $\sqrt{2}/4 \pi$. 
Experimentally, the most favorable situation is when one of the photons
is real or almost real: $q_1^2 \sim 0$.
In this case, we will denote the form factor by 
$ F_{\gamma^*\gamma \pi^0}(Q^2)$, where  $Q^2 \equiv -q_2^2$
is the  virtuality of the other photon. It 
should be sufficiently large for pQCD 
to be applicable.
In general,  a  power-like behavior 
of $F_{\gamma^*\gamma \pi^0}(Q^2)$ in the large-$Q^2$ 
limit can be generated by three basic regimes
(see Fig.1).

\begin{figure}[htb]
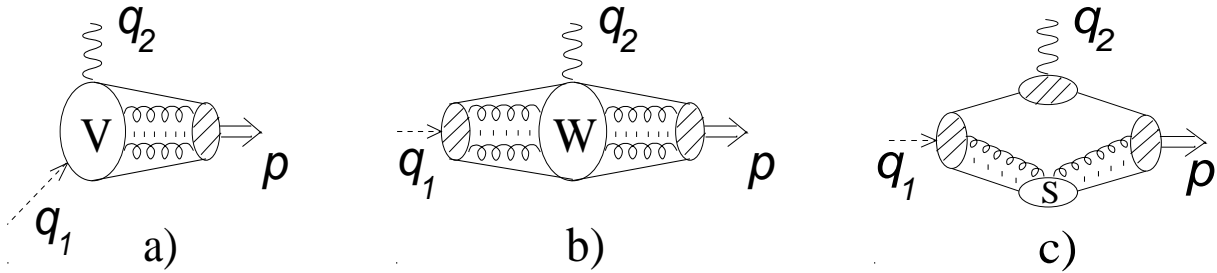

\mbox{
   \epsfxsize=3.7cm
 \epsfysize=3.5cm
 \hspace{0cm}  
  \epsffile{fig1a.eps}  } \hspace{1cm}
\mbox{
   \epsfxsize=5cm
 \epsfysize=3.5cm
 \hspace{0cm}  
  \epsffile{fig1b.eps}  }\hspace{1cm}
\mbox{
   \epsfxsize=4.5cm
 \epsfysize=3.5cm
 \hspace{0cm}  
  \epsffile{fig1c.eps}  }
  \vspace{0.5cm}
{\caption{\label{fig1}
Structure of factorization for the $F_{\gamma^*\gamma \pi^0}(Q^2)$ 
form factor at large $Q^2$. 
   }}
\end{figure}

The dominant contribution
is provided by the   first regime (Fig.$1a$) 
which corresponds to  large virtuality flow
through a subgraph $V$ containing both photon vertices.
 The power counting estimate 
for the large-$Q^2$ behavior 
of  such a configuration  with arbitrary number of external lines of $V$
is given by (see refs.\cite{pl,rr})
\begin{equation} 
F(Q^2)  \lesssim Q^{-\sum \limits_i t_i}
\label{2}    \end{equation}
where $t_i$'s are  twists (dimension minus spin) of the 
quark and gluon external lines of $V$,
with $t=1$ for the quarks and $t=0$ for the gluons
in a covariant gauge.  Hence,  for the leading term, 
one should take  the minimal number of quark lines
(two in our case)  while the number of the 
gluonic $A$-fields is arbitrary.  
Generically, the leading  contribution of this type  can be written as  
\begin{equation} 
\ F_{\gamma^*\gamma^*  \pi^0}(q_1, q_2) = 
\int   C(\xi, \eta, q_1, q_2 ; \mu^2)  \,
\langle p | {\cal O} (\xi, \eta) | 0 \rangle  |_{\mu^2} 
 d^4 \xi d^4 \eta \,  ,
\label{1}  \end{equation}
where the parameter $\mu^2$ is  the factorization scale,
$C(\xi, \eta, q_1, q_2)$ corresponds to the short-distance
amplitude  with two external quark 
lines  and ${\cal O}(\xi , \eta)$ is  a composite
operator ${\cal O}(\xi , \eta) \sim \bar q (\xi) \gamma_5 \gamma_{\nu}
E(\xi, \eta;A) q(\eta)$.  
The  path-ordered exponential 
$$E(\xi, \eta;A)\equiv P \exp \left 
( ig \int_{\eta}^{\xi} A_{\mu}(z) dz^{\mu} \right )$$ 
 of the gluonic field $A$ results from summation
over   external gluon lines of $V$. 
For the quark propagator, $e.g.,$
one has
\begin{equation} 
S^c(\xi - \eta) + \int S^c(\xi - z) \gamma^{\mu} g A_{\mu}(z) S^c(z - \eta) \, d^4z
+ \ldots = E(\xi, \eta;A) \, S^c(\xi - \eta) \bigl [ 1 + O(G) \bigr ]
\label{1E}  \end{equation}
where $O(G)$ depends on the gluonic fields 
through the gluon field strength tensor
$G_{\mu \nu}$ and its covariant  derivatives.
Since  $G_{\mu \nu}$ is asymmetric with respect to 
the interchange of the indices $\mu$, $ \nu$, 
it  should  be treated as  a twist-1 field.

Basically,  the contribution (\ref{1})  
is analogous to 
 the quark-antiquark term  of the  standard 
operator product expansion 
for  $J^{\alpha }(0) J^{\beta }(z)$.
In this form,   the operator ${\cal O}(\xi , \eta)$ 
still contains non-leading twist terms.
To get the lowest-twist part, we should expand
${\cal O}(\xi , \eta)$
into the   Taylor series 
\begin{equation} 
\bar q (\xi) \gamma_5 \gamma_{\nu}
E(\xi, \eta;A) q(\eta) = \sum \limits_{n=0}^{\infty} 
\frac1{n!} \Delta^{\nu_1} \Delta^{\nu_2} 
\ldots \Delta^{\nu_n}  
\bar q (\xi) \gamma_5 \gamma_{\nu} D_{\nu_1}D_{\nu_2}
\ldots D_{\nu_n} q(\xi) \quad ;  \quad   \Delta = \eta - \xi
\label{3}  \end{equation} 
and pick out only the symmetric-traceless part 
$\bar q  \gamma_5 \{ \gamma_{\nu} D_{\nu_1}D_{\nu_2}
\ldots D_{\nu_n} \} q $ 
of each local operator from this expansion.
The traces correspond to  operators with contracted 
covariant derivatives $D^{\nu}D_{\nu}$  
 which, for
dimensional reasons, are accompanied
by powers of the interval $(\xi - \eta)^2$. 
Likewise, the $(\xi - \eta)^2$ factors 
 produce  extra powers of $z^2$ 
after integration over $\xi$ and $\eta$. 
Finally, each power of $z^2$ results in an extra power of $1/Q^2$,
$i.e.,$ each pair of contracted covariant derivatives 
$D^{\nu} \ldots D_{\nu}$ in a 
higher-twist operator produces 
$1/Q^2$ suppression at large $Q^2$. 
Hence,  the twist-2 part of ${\cal O}(\xi , \eta)$ 
corresponds to the
lowest term of  the  expansion over $(\xi - \eta)^2$ 
\begin{equation} 
\left. {\cal O}(\xi , \eta) = {\cal O}(\xi , \eta) \right 
|_{(\xi - \eta)^2=0} + O((\xi - \eta)^2).
  \label{4} \end{equation}
The  light-cone  matrix element
can be parametrized in terms of the pion distribution amplitude (DA)
$\varphi_{\pi}(x)$:
\begin{equation} 
\left. \langle 0| {\cal O}_{\nu}(\xi , \eta)
| {\pi^0},{p} \rangle \right |_{(\xi - \eta)^2=0} 
= i p_{\nu} \int_0^1  e^{- i x (\xi p) - i \bar x (\eta p)} 
\varphi_{\pi} (x) dx  \  ,
\label{5}    \end{equation}
which gives the probability amplitude that the fast-moving pion 
is a    $\bar q q$ pair with its longitudinal momentum $p$ 
shared among the quarks in fractions $x$ and  $\bar x \equiv (1-x)$
(throughout the paper, we use the ``bar'' convention
for the momentum fractions:
$ \bar x \equiv 1-x, \bar y \equiv 1-y$, $etc.$).  
Substituting this representation into the
generic expression (\ref{1}), we obtain the hard scattering formula
\begin{equation} 
  F_{\gamma^*\gamma \pi^0}(q_1,q_2) =
  \frac{4 \pi}{3} \int_0^1 T(q_1,q_2; xp, \bar x p) \,
\varphi_{\pi}(x) \, dx  \  ,
\label{6}    \end{equation}
where the factor $4 \pi/3$ is due to
our normalization of the form factor
and  $T(q_1,q_2; k, \bar k)$ is the amplitude for the subprocess
$\gamma (q_1) \gamma^*(q_2) \to \bar q (\bar k) q (k)$. 
Calculating this lowest-twist amplitude
in the momentum representation, we should realize that 
the neglect of the higher-twist operators 
having extra  $D^2$ is equivalent to taking   
$k^2=0, \bar k^2 =0$ 
for the  external quark momenta.   
In general, this limit is singular for diagrams with loops,
and one should  regulate
the resulting mass singularities $\ln k^2$  in some way,
$e.g.,$ by dimensional regularization or by taking massive quarks 
and  $k^2=m_q^2$. In the latter case, only the logarithmic
$m_q$-dependence should be kept in the final result:  
keeping  the power terms $m_q^2/Q^2$   exceeds,
for light quarks,  the accuracy 
of  the method.  The  subsequent  procedure is to
split the  logarithms  $\ln (Q^2/m^2) $  into the long-distance 
and short-distance parts $\ln (Q^2/m^2)  = \ln (Q^2/\mu^2) +\ln (\mu^2 /m^2) $
and absorb the long-distance ones $\ln (\mu^2 /m^2) $ into the
pion distribution amplitude: $\varphi_\pi (x) \to \varphi_\pi (x; \mu)$.

\begin{figure}[htb]
\hspace{5cm}
\mbox{
   \epsfxsize=5cm
 \epsfysize=3.5cm
 \hspace{0cm}  
  \epsffile{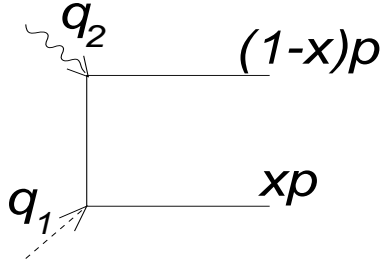}  } 
  \vspace{0.5cm}
{\caption{\label{fig0}
Lowest-order diagram. 
   }}
\end{figure}

Thus, the lowest-twist contribution 
corresponds to the  parton picture in which  
only the longitudinal (proportional to $p$)
components of the external quark momenta appear. 
In the lowest order (see Fig.2),  the amplitude for transition 
of two photons into the quark-antiquark pair 
with collinear lightlike momenta $xp$,
$\bar x p$ 
is given\footnote{In fact, there are two
diagrams obtained from one another by the interchange 
of photon vertices. However, due to the 
symmetry of the distribution amplitude
$\varphi_{\pi}(x) = \varphi_{\pi}(1-x)$,
their contributions can be united.}
by the quark propagator: 
 \begin{equation}  
T_0 (x,Q^2)  =   \frac1{-(q_1 - xp)^2} = \frac1{xQ^2}  \,  .
  \label{7} \end{equation}
and  the pQCD result \cite{bl80}
 for the large-$Q^2$ behavior of the
form factor is
\begin{equation}  
F_{\gamma^* \gamma\pi} (Q^2) = \frac{4 \pi}{3} \int_0^1 
 \frac{\varphi_{\pi}(x)}{xQ^2} \, dx \,
\equiv \frac{4\pi f_{\pi} }{3Q^2} I_0 .
 \label{8} \end{equation}
 Necessary 
nonperturbative information  
is accumulated   in 
the same integral 
 \begin{equation} 
I_0 = \frac1{f_{\pi}} \int_0^1 
 \frac{\varphi_{\pi}(x)}{x} \, dx =
\frac{Q^2}{f_{\pi}} \int_0^1 T_0(x,Q^2) \,  \varphi_{\pi}(x) \, dx 
\label{8A} \end{equation}
 that appears  in the one-gluon-exchange 
diagram  for the  pion electromagnetic  form factor  
\cite{pl80,blpi79,czas}.
The  value of $I$ depends on the shape of the 
pion distribution amplitude $\varphi_{\pi}(x)$.
In particular,  using  the 
asymptotic form \cite{pl80,blpi79} 
\begin{equation} 
\varphi_{\pi}^{as}(x) = 6 f_{\pi} x (1- x )  
\label{AS} \end{equation} 
 gives $I_0^{as}=3$.  If one takes the
Chernyak-Zhitnitsky ansatz  \cite{cz82}
\begin{equation}
\varphi_{\pi}^{CZ}(x) = 30 f_{\pi} x(1-x)(1-2x)^2  \,  , 
\label{CZ} \end{equation}
the integral $I_0$ increases by a sizable factor of 5/3:
$I_0^{CZ}=5$ 
and one can hope that this difference can be used for 
an experimental
discrimination between the two competing models for the pion DA.

Since one of the photons has a small virtuality,
one should, in principle, also 
take into account the regime (see Fig.$1b$)
involving a long-distance propagation  in the 
$q_1$-channel,
with large momentum flowing through a central subgraph $W$
containing  only the virtual photon vertex. 
In the lowest order, this subgraph corresponds to 
a hard-gluon exchange, just like in the asymptotically 
leading pQCD contribution to the pion electromagnetic form factor.
The power counting for such a contribution
into $F_{\gamma^*\gamma \pi^0}(Q^2)$ is given by 
\begin{equation} 
F(Q^2)  \lesssim Q^{- t_{{\cal O}_1} - t_{{\cal O}_2}} \, , 
\label{2A}    \end{equation}
where $t_{{\cal O}_i}$, $i=1,2$  are the twists of composite 
operators ${\cal O}_i$  corresponding to $q_1$- and $p$-channel, 
respectively. Taking into account that   twist of a gauge-invariant
color-singlet composite operator ${\cal O}_i$ 
cannot be less than 2, we conclude that  this regime gives a nonleading 
$O(1/Q^4)$ contribution. 

The third regime (Fig.$1c$) corresponds to Feynman
mechanism, $i.e.$ to  a situation when the passive quark 
is soft. Using the wave function terminology, we can say
that $F_{\gamma^*\gamma \pi^0}(Q^2)$ in this regime 
is given by an overlap of soft wave functions
describing  the initial and final state.
This contribution also behaves like $1/Q^4$ 
at large $Q^2$.

\subsection{Handbag diagram  and transverse momentum}

For the  OPE contribution, the simplest  power corrections 
come either from the traces of the two-body operator 
${\cal O}(x,y)$  which appears in  the handbag diagram
or from a direct insertion of gluon lines with physical polarizations
into the propagator connecting the photon vertices.
Since $D^{\nu}D_{\nu}$ can be interpreted in the 
momentum representation  as the 
(generalized) virtuality $k^2$  of the quark field, 
the higher-twist operators containing  $D^{\nu}D_{\nu}$
look like a  natural  candidate for description of 
 the effects due to the 
transverse momentum  of the quarks. 
However,  there are some 
practically important 
amplitudes which,
due to their simple singularity structure,
 are ``protected''  from the  towers of 
$(D^2)^n$-type 
higher-twist corrections. 
The most well-known example  is given by 
the classic ``handbag'' diagram for deep inelastic 
scattering.    The lowest-order diagram 
for the $\gamma^*\gamma \to \pi^0$
form factor (Fig.2) has similar properties.
Consider its  analog in a toy scalar model
\begin{equation}
F(q_2,p) = \frac1{4 \pi^2} \int  e^{-iq_2z}  \langle 0 | 
\phi (0) \phi (z) |p \rangle \frac{d^4z}{z^2} \, . 
\label{9} \end{equation}
The first term in the $z^2$-expansion for  the matrix element  
\begin{equation}
 \langle  0 | \phi (0)  \phi (z) | p \rangle
=  \xi_2(zp) + z^2 \xi_4(zp) + (z^2)^2 \xi_6 (zp) + \ldots
  \label{10} \end{equation}
corresponds to the twist-2 distribution amplitude 
while subsequent  terms correspond to operators containing an 
increasing number of $\partial^2$'s. 
It is straightforward to observe that,
while the twist-2 term produces the $1/Q^2$ contribution,
the twist-4 term 
is accompanied by an  extra $z^2$-factor which 
completely kills the $1/z^2$-singularity of the quark propagator,
and  $d^4z$ 
integration gives $\delta^4(q-xp)$, which is  
 invisible for large $Q^2$. The same is evidently true for all the terms
accompanied by higher powers of $z^2$. 
This means that the handbag diagram contains only one  
term with a powerlike behavior for large $Q^2$:  it cannot generate  
higher powers of $1/Q^2$  which one could interpret as 
the $(\langle k^2 \rangle/Q^2)^n$  expansion. 
Since only the  $z^2 =0$ projection  of the  bilocal operator 
survives,   we can parametrize  
\begin{equation}
 \langle  0 | \phi (0)  \phi (z) | p \rangle = 
\int_0^1 \varphi (x) e^{-i \bar x (zp)} \, dx +\ldots  \ , 
  \label{11} \end{equation}
where  the dots stand for terms producing the ``invisible''
contributions, 
and write the  lowest-order term as
 \begin{equation}
F(q_2,p) = 
- \int_0^1 \frac{\varphi (x)}{ (q_2 - \bar x p)^2 }  \, dx \,  =
- \int_0^1 \frac{\varphi (x)}{ (q_1-  x p)^2 }  \, dx \,  =
 \int_0^1 \frac{   \varphi (x) } {xQ^2 +x \bar x p^2} \, dx .
\label{12}  \end{equation}
Hence, the handbag contribution in this case 
contains only  the hadron-mass corrections (cf.\cite{ht}),
but  it gives   no information about finite-size effects. 
In the momentum representation,
the origin of this phenomenon can be traced
to the fact that a straightforward  expansion 
of the propagator is just  in terms of  traceless
combinations:
\begin{equation}
 \frac{1}{(q-k)^2}  = 
\theta (|k| <|q|) \sum_{n=0}^{\infty}  \frac{2^n}{(q^2)^{n+1}}
q^{\mu_1} \ldots q^{\mu_n} \{k_{\mu_1} \ldots k_{\mu_n} \}
+ \theta (|k| > |q|) \sum_{n=0}^{\infty}  \frac{2^n}{(k^2)^{n+1}}
q^{\mu_1} \ldots q^{\mu_n} \{k_{\mu_1} \ldots k_{\mu_n} \} .
\label{13}   \end{equation}
The handbag contribution corresponds to $|k| <|q|$,
and this part of Eq.(\ref{13})  without any approximation
 produces an  expression equivalent to 
treating the $k$-momentum as purely longitudinal $k= \bar x p$.

It is worth noting here that though the hadron-mass 
corrections have a $powerlike$ behavior  $(p^2/Q^2)^n$, 
they should not be classified as $higher$-$twist$ corrections:
they result from the kinematic hadron-mass dependence 
of the $lowest$-$twist$ contribution.
For deep inelastic scattering, the possibility to
calculate the target-mass corrections 
within the lowest-twist contribution
is known as the $\xi$-scaling phenomenon \cite{Nacht,gp}.
As emphasized by K.Ellis {\it et al.} \cite{ht},
the $\xi$-scaling phenomenon can be also understood 
in terms of the primordial transverse momentum,
if one takes into account that,  
for the lowest-twist term, the  transverse 
momentum distribution is totally due to the 
non-zero hadron mass, $i.e.,$ it has a purely kinematic nature
and for this reason can be calculated exactly.
The quark propagator in QCD has a stronger singularity 
$\hat z /z^4$. As a result, the handbag-type  contribution in QCD 
 contains  a twist-4 operator with extra $D^2$ \cite{gorskij},
but no operators with higher powers of $D^2$.

One may argue that there is another part 
in Eq.(\ref{13}),  when  $k$ is large ($i.e.,$ $|k| > |q|$).
In this case, 
 the $k$-line corresponds to high virtualities.
If such a large momentum goes directly into the soft 
hadronic wave function, the $Q^2$-behavior of
 such a  contribution  repeats the $k^2$-dependence 
of the soft wave function,
$i.e.,$ very rapidly (say, exponentially) 
 decreases  with  $Q^2$ (see Section IV$C$ 
below for an explicit 
illustration).
A more favorable possibility is  
when the large momentum  by-passes the wave function. 
Such a configuration can give a leading-power contribution.  
In the latter case, 
the large virtuality flows through several
lines forming a subgraph with the same (minimal 
 possible)  number of external quark lines as the 
lowest-order leading twist contribution.
In the QCD factorization scheme, 
the relevant contribution  produces 
a part  of a higher-order 
coefficient function (see Fig. 1$a$).

\subsection{One-loop radiative correction to the coefficient function}

At one loop, the coefficient function for the $\gamma^*\gamma \to \pi^0$
form factor was calculated in refs.\cite{AuCh81,braaten,kmr}:
\begin{equation}
T(x, Q^2;\mu^2) = \frac{1}{ xQ^2} \biggl 
\{ 1+ C_F {{\alpha_s}\over{2 \pi}} \biggl 
[\biggl (\frac3{2} + \ln x \biggr )  \ln (Q^2 / \mu^2) + 
\frac1{2} \ln^2 x - \frac{x \ln x}{2(1-x)}  - 
\frac9{2} \  \biggr ] \biggr \} . \label{14}
 \end{equation}

\begin{figure}[htb]
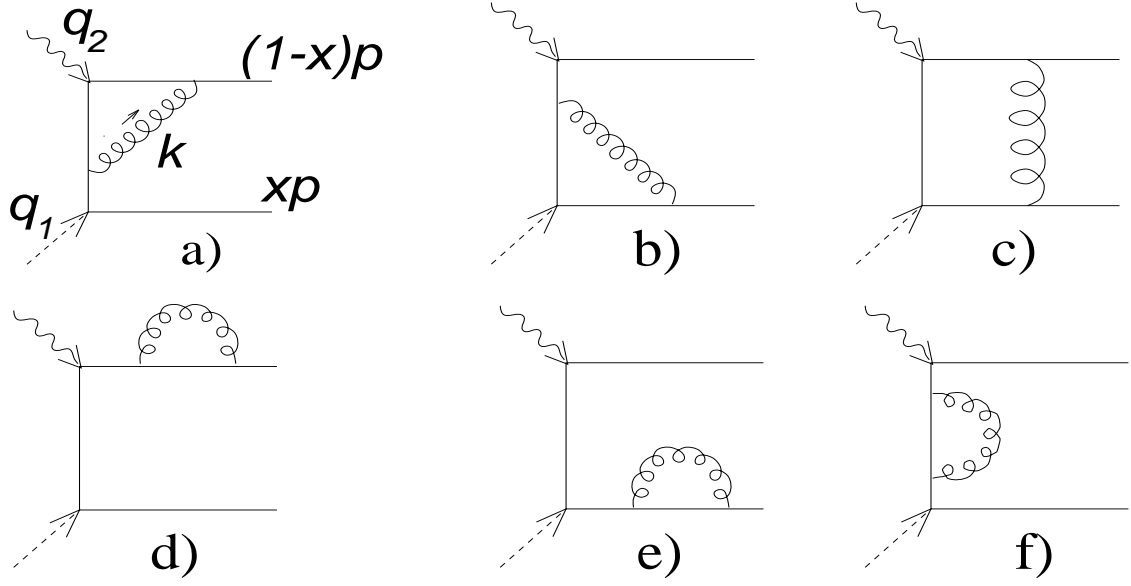

\mbox{
   \epsfxsize=5cm
 \epsfysize=3.5cm
 \hspace{0cm}  
  \epsffile{fig2a.eps}  } \hspace{1cm}
\mbox{
   \epsfxsize=3.5cm
 \epsfysize=3.5cm
 \hspace{0cm}  
  \epsffile{fig2b.eps}  }\hspace{1cm}
\mbox{
   \epsfxsize=3.5cm
 \epsfysize=3.5cm
 \hspace{0cm}  
  \epsffile{fig2c.eps}  }

  \vspace{0.5cm}
\hspace{0cm}
\mbox{
   \epsfxsize=3.5cm
 \epsfysize=3.5cm
 \hspace{0cm}  
  \epsffile{fig2d.eps}  } \hspace{2.5cm}
\mbox{
   \epsfxsize=3.5cm
 \epsfysize=3.5cm
 \hspace{0cm}  
  \epsffile{fig2e.eps}  }\hspace{1cm}
\mbox{
   \epsfxsize=3.5cm
 \epsfysize=3.5cm
 \hspace{0cm}  
  \epsffile{fig2f.eps}  }
  \vspace{0.5cm}
{\caption{\label{fig2}
One-loop diagrams. 
   }}
\end{figure}

In full compliance with the factorization theorems \cite{tmf78,bl80}
(see also \cite{20,facter,facter2}),  
the one-loop contribution contains no 
Sudakov double logarithms $\ln^2 Q^2$ of the
large momentum transfer $Q$. 
Physically, this result is due to the color neutrality of the pion.
In the axial gauge, the Sudakov double logarithms
appear in the box diagram $3c$ but they are cancelled 
by similar terms from  the quark self-energy corrections $3d,e$.
In Feynman gauge, the double  logarithms $\ln ^2 Q^2$
simply do not appear in any one-loop diagram.
 It is easy to check that the term containing 
 the    logarithm $\ln (Q^2 / \mu^2)$  
has  the form of  convolution
\begin{equation}
\frac{1}{x Q^2} \, C_F {{\alpha_s}\over{2 \pi}} 
\biggl (\frac3{2} + \ln x \biggr )  =
\int \limits_0^1 \frac{1}{\xi Q^2} \, V(\xi, x)  \, d\xi 
\label{15} \end{equation}
of the lowest-order (``Born'') term $T_0(\xi, Q^2) = 1/\xi Q^2$
and the  kernel 
\begin{equation}
V(\xi,x) =  {{\alpha_s}\over{2 \pi}}\, C_F \, 
\left [ { \xi \over  x} \, 
\theta(\xi <  x) \left ( 1 + {1 \over x- \xi } \right )
+ {\bar \xi \over \bar x} \,  \theta(\xi > x) 
\left ( 1 + {1 \over \xi - x} \right )
\right ]_+ \label{16}
 \end{equation}
governing the evolution of the pion distribution amplitude.
The ``+''-operation is defined here, as usual \cite{altpar},  by 
\begin{equation}
[F(\xi,x)]_+ = F(\xi,x) -\delta (\xi-x) \int \limits_0^1
 F(\zeta, x)\, d \zeta \,  . \label{17}
 \end{equation}

Since the asymptotic   distribution amplitude
is the eigenfunction of the evolution kernel 
$V(\xi,x)$  corresponding to zero eigenvalue,
\begin{equation}
\int_0^1 \, V(\xi,x) \, \varphi^{as}(x) \, dx = 0 \, , 
\label{18} \end{equation}  
the coefficient $\frac32 + \ln x$ of the $\ln (Q^2/\mu^2)$ term 
vanishes after the $x$-integration with $\varphi^{as}(x)$.
Hence,  the size of the one-loop correction for the 
asymptotic  DA is $\mu$-independent and determined 
by the remaining terms.  The  $I$-integral
 \begin{equation} 
I \equiv 
\frac{Q^2}{f_{\pi}} \int_0^1 T(x,Q^2) \,  \varphi_{\pi}(x) \, dx 
\label{8I} \end{equation}
(cf. Eq.(\ref{8A}))   
then can be written as 
\begin{equation}
 I \, |_{\varphi =\varphi^{as}}  =
 3 \left\{ 1 - \frac{5}{2} \,  C_F {{\alpha_s}\over{ 2\pi}}\right \}. 
\label{19}  \end{equation}
The negative coefficient $-5/2$ here comes from the 
constant term  $-9/2$ (see Eq.(\ref{14}))  partially compensated
by two logarithmic terms which give together $+2$,
with $+7/4$ generated by the 
 $\frac1{2} \ln^2 x$ contribution  and $+1/4$ by $-x \ln x /[2(1-x)]$ term.
With $C_F=4/3$, the net factor  is $[1- \frac5{3} \alpha_s/ \pi]$.
Hence,  for $\alpha_s/ \pi \approx 0.1$, the one-loop correction  
is less than $20 \%$ and    the $\alpha_s/ \pi$ expansion 
looks  ``reasonably convergent''.
Taking the CZ form for $\varphi(x;\mu)$,
we  get 
\begin{equation}
  I \, |_{\varphi(x,\mu) =\varphi^{CZ}(x)} \,  =
 5 \left\{ 1 - C_F {{\alpha_s}\over{ 2\pi}} \left 
( \frac5{6} \ln (Q^2/\mu^2) + \frac{49}{72} \right ) \right\}.
 \label{20} \end{equation}
Again, the negative coefficient $-49/72$
comes from the $-9/2$ term 
compensated by an increased contribution from the logarithmic terms:
 $\frac1{2} \ln^2 x$   gives $+263/72$   
and $-x \ln x /[2(1-x)]$ gives $1/6$.
For $\mu =Q$, the one-loop modified factor is 
$[1- \frac{49}{108} (\alpha_s/ \pi)]$,
$i.e.,$   the total correction is  smaller 
than that for the asymptotic DA.
Since the  result  is $\mu$-dependent in this case,
by an  appropriate choice 
of $\mu$, namely, taking $\mu =  e^{\frac{49}{120}}Q
\approx 1.5 \, Q$  we can formally get a vanishing 
$ {\cal O} (\alpha_s)$ correction.
Then the one-loop expression for the form factor 
would coincide with the lowest-order formula, but 
with the distribution amplitude $\varphi_{\pi}^{CZ}(x;\mu)$
evolved to the scale $\mu \approx 1.5 \, Q$.
However, at this scale, $\varphi_{\pi}(x;\mu)$
does not necessarily have the CZ form.
To treat the evolution in a consistent way, 
we set the boundary condition that
$\varphi_{\pi}^{CZ}(x;\mu)$  has the canonical CZ form 
$\varphi_{\pi}^{CZ}(x) \equiv 30 \, f_{\pi} \, x \bar x \, (1-2x)^2$
at some specific scale $\mu =Q_0$ (the original derivation
\cite{cz82} assumes $Q_0=0.5 \, GeV$).  Taking into account
that $\varphi_{\pi}^{CZ}(x)$ is a combination of 
two lowest eigenfunctions of the evolution kernel, we can write 
the solution of the evolution equation in the leading
logarithm approximation:
\begin{equation}
\varphi_{\pi}^{CZ}(x;\mu) = \varphi_{\pi}^{as}(x) + 
\{\varphi_{\pi}^{CZ}(x) - \varphi_{\pi}^{as}(x)\} \left [\frac{\ln Q_0^2/ \Lambda^2}{
\ln \mu^2/ \Lambda^2 } \right ]^{\gamma_2/\beta_0}\, , 
\label{21} \end{equation} 
where $\gamma_2 = 50/9$ is the relevant anomalous
dimension and $\beta_0 =11-\frac23 N_f$ is the
lowest coefficient of the QCD $\beta$-function.
In what follows, we take $N_f=3$ and $\beta_0=9$. 
 Choosing $\mu =Q$, we get for the $I$-integral  (cf. \cite{kroll2})
\begin{equation}
  I \, |_{\varphi_{\pi}(x,Q_0) =\varphi_{\pi}^{CZ}(x)} \,  =
 3\left\{ 1 -\frac53 {{\alpha_s}\over{\pi}} \right \}
\left ( 1 - \left [\frac{\ln Q_0^2/ \Lambda^2}{
\ln Q^2/ \Lambda^2 }  \right ]^{50/81}  \right )
+5 \left\{ 1 -\frac{49}{108} {{\alpha_s}\over{\pi}} \right \} 
\left [\frac{\ln Q_0^2/ \Lambda^2}{
\ln Q^2/ \Lambda^2 }   \right ]^{50/81} \, . 
 \label{22} \end{equation}

Note that the  $\ln^2 x$  term generates a   
larger positive contribution 
 for $\varphi_{\pi}^{CZ}(x)$
 because $\varphi_{\pi}^{CZ}(x)$ is more concentrated 
in  the end-point region $x \sim 0$ than $\varphi_{\pi}^{as}(x)$. 
Furthermore,  if the distribution amplitude  
is  extremely  concentrated in  
the end-point region $x \sim 0$,
a  positive contribution from the  $\frac1{2} \ln^2 x$ 
term   dominates the correction and   generates 
a large positive  net effect. In such a situation,
the  one-loop correction vanishes only if 
$\mu =a Q$ with $a<1$. The broader the DA, 
the smaller should be the  parameter $a$  
 which reduces the one-loop expression to the lowest-order
one.  
Since the effective 
normalization scale is smaller for a 
broader DA, 
perturbative QCD  applicability is postponed to higher $Q^2$.  
One may speculate that this phenomenon simply 
indicates that for a broad DA the quark 
virtuality $xQ^2$ is a more natural 
choice for the  effective  factorization scale
 than  the 
photon virtuality $Q^2$ ($i.e.,$ $a\sim \langle x \rangle$)
 and   pQCD  is applicable
only if the average $xQ^2$ rather than 
$Q^2$ itself is large enough.
One faces a similar situation studying the 
pQCD contribution to  the pion form factor.
The average virtuality $\langle xyQ^2 \rangle $ of the exchanged gluon
 in that case is essentially smaller than $Q^2$ and one may question
both the {\it self-consistency } and {\it reliability}
 of the pQCD analysis at accessible energies
\cite{ils,rad}. 
In ref.\cite{listerman}, it was argued that due to the
Sudakov effects in the impact parameter space, 
 the pQCD treatment of the lowest-twist one-gluon-exchange
term  for the pion form factor is 
{\it self-consistent}\footnote{Note, that  self-consistency
of the pQCD expansion (small $\alpha_s$ corrections)
for the lowest-twist  term does not necessarily
mean that pQCD is reliable,
since power corrections $(M^2/Q^2)^n$ 
can still be large (see discussion
at the end of Section V).}
at smaller
$Q^2$ than  suggested by the  estimates of 
the  magnitude of the average  gluon virtuality $xyQ^2$.
One may  expect that  similar
effects manifest themselves also in   the $\gamma^* \gamma \pi^0$ 
form factor.  Indeed, our numerical  analysis of
the one-loop correction  shows that taking $a=1$
(rather than $a= \langle x \rangle$)  provides   
a good  choice for  the factorization scale.
It is accompanied by  a small one-loop correction 
even for a broad DA of CZ type.

It is worth noting here that,
even without incorporating
the impact parameter representation, 
one can observe  some traces of the Sudakov effects 
 in the structure of the 
one-loop coefficient function in the region of 
 small fractions $x$. 
As explained earlier, the one-loop term is  obtained by calculating 
the $\gamma^*\gamma \to \bar q q$ amplitude for massive on-shell
quarks  with subsequent absorption of the mass logarithms in the form 
$\ln (\mu^2 /m^2)$ into the distribution amplitude. 
When the virtuality $xQ^2$ of the
quark line connecting the photon vertices
becomes small, the vertex correction for the virtual photon (Fig.$3a$) 
is dominated (in Feynman gauge) by the {\em off-shell} 
Sudakov double logarithm 
which can be written as $$- {{\alpha_s}\over{2 \pi }}
\,  C_F \ln \frac{Q^2}{m^2} \ln \frac{Q^2}{xQ^2}\, $$
where $xQ^2$ is the virtuality of the hard quark.
Of course, since this virtuality is 
parametrically of the order of $Q^2$,
we get  only a single logarithm with respect to $Q^2$, namely, 
$({\alpha_s}/{2 \pi})\,  C_F \ln (Q^2/m^2) \ln x$
(cf.(\ref{14})), just as required for  factorization.
However, if we  write the sum of 
two terms 
$${{\alpha_s}\over{4 \pi }} \, C_F \left [ \ln^2 x + 2 
\ln  \frac{Q^2}{m^2}\ln x \right ]$$ 
which dominate  the small-$x$ region 
as  $${{\alpha_s}\over{4 \pi }} \, C_F
\left [ \ln^2 \frac{x Q^2}{m^2}  - \ln^2 \frac{Q^2}{m^2} \right ] \, , $$ 
we see that it converts into  
the standard {\em on-shell}  
Sudakov double logarithm $$ - \frac{\alpha_s}
{4 \pi } \, C_F \ln^2  \frac{ Q^2}{m^2} $$ when $xQ^2 \sim m^2$.
Of course, the  region  
where $xQ^2$ is parametrically
of the order of the $IR$ cut-off $m^2$ 
is outside the formal applicability region
of the factorization approach, and there is no surprise that
double logarithms 
of $Q^2$   appear there. 
Note the well-known difference $\alpha_s/2\pi \to \alpha_s/4\pi$
between the {\it off}- and {\it on-shell} forms
of the double logarithms. 
In higher orders, Sudakov  logarithms are expected to 
exponentiate producing the 
Sudakov form factor\footnote{For the pion EM form factor, 
  exponentiation of a similar combination 
$( C_F \alpha_s/4 \pi )
 [ \ln^2 (x y Q^2 / m^2)  - \ln^2 (Q^2 / m^2)  ]$
suggested in ref.\cite{ditrad} 
was verified by a two-loop calculation \cite{ohrndorf}.}
$\exp[-(\alpha_s/4 \pi) \, 
 C_F\ln^2 ( Q^2/m^2)]$, 
and  the region of very small  $xQ^2$ is 
relatively suppressed due to Sudakov effects.

This also means that  taking 
$\mu^2 \sim  xQ^2$ in Eq.(\ref{14}) 
is not an optimal choice,
 since  it 
is accompanied by   a negative rather than vanishing  correction. 
Indeed, the original motivation
to take a lower scale $\mu <Q$ was to compensate
the positive contribution 
from the $\ln^2 x$ term. However, 
 taking  $\mu^2 \sim  xQ^2$ in Eq.(\ref{14})
for a wide DA generates a negative   $(- \ln^2 x)$
 term which over-kills the original 
positive  $\frac1{2} \ln^2 x$ term 
and converts its sign in the net result.
A negative correction, in its turn,  
 suggests that a  larger factorization scale 
is  a better choice.
This indicates that, for a broad DA, the    typical  
distances probed  in the hard subprocess 
are larger than those corresponding to $1/Q^2$
but  smaller than  those corresponding to 
the inverse of the average quark virtuality $xQ^2$.

As we will see in the next section, the 
modified factorization \cite{listerman} is similar 
to the choice $\mu^2 \sim  xQ^2$
and for this reason it is accompanied by a
negative correction.  We will also 
explicitly show that the latter,
in full accordance with the MFA analysis \cite{bottssterman},
 can be explained by  
Sudakov effects in the impact parameter space.

\section{One-loop radiative corrections and transverse momentum}

\subsection{Vertex correction for virtual photon and Sudakov effects}

To establish the connection between standard 
and modified factorization approaches, 
we give below a rather  detailed 
discussion  of  the structure of the 
one-loop coefficient function 
using  the Sudakov decomposition for  the
loop momenta. We  use  the same definition
of transverse momentum $k_{\perp}$ as in 
ref.\cite{bottssterman,listerman}, introduce
 the  impact parameter $b_{\perp}$
and then translate our results into the  $b_{\perp}$-space. 
To be able to make a diagram by diagram comparison 
with ref.\cite{braaten},  we  use  Feynman gauge. 
This also allows us to give an independent 
one-loop derivation  of the $b_{\perp}$-space Sudakov 
effects which complements 
the general approach   \cite{bottssterman}
based on the analysis in   the  axial gauge\footnote{
In a recent paper\cite{licov}, Li gave a covariant gauge 
derivation of the 
modified factorization  for inclusive processes
and heavy-quark decays. However, in technical
implementation, his approach is quite different from ours.}.
We find it also instructive to demonstrate
how the  $b_{\perp}$-space double logarithms 
appear in a situation in which 
double logarithms of $Q^2$ are absent in any diagram.

We start with the diagram $3a$ which is the   most natural  suspect
in a search for Sudakov effects in Feynman gauge.
According to general rules, calculating the
 coefficient function one  should  assume that 
external quarks carry purely longitudinal  lightlike 
momenta $xp$ and $\bar xp$.
Using $p$ and $q_1$ (abbreviated in this section to
$q$ for convenience) as the basic Sudakov
light-cone variables, we write the 
momentum $k$ of the  emitted gluon  as
\begin{equation} 
k=  (\xi - x) p + \eta q + k_{\perp}
\label{24}    \end{equation}
and then take  the $\eta$-integral by residue. 
After that, the  contribution
of Fig.$3a$ (and any other one-loop diagram)
can be schematically written as 
\begin{equation} 
T^{(1)}_i(x,Q^2) =   
 \frac{\alpha_s}{2 \pi} \,  C_F 
\int_0^1 d \xi \int M_i(x,Q^2;\xi, k_{\perp}) \, 
\frac{d^2 k_{\perp}}{2 \pi} \, 
\equiv \frac{\alpha_s}{2 \pi} \,  C_F \, t_i(x,Q^2) . 
  \label{25} \end{equation}
The internal amplitude $M_a (x,Q^2;\xi, k_{\perp})$ 
for the diagram $3a$
is given by 
\begin{equation}  
M_a (x,Q^2;\xi, k_{\perp})    ={1\over {xQ^2} } \left \{
           -  \left ( {\bar \xi \over \bar x} \right ) 
              { Q^2 + k_{\perp}^2/\bar \xi
                 \over k_{\perp}^2 \left [\xi  Q^2 + k_{\perp}^2
 /\bar \xi\right ]
                           }  
\, \theta (\xi>x)  
 + {  k_{\perp}^2   \,  \theta (\xi<x)  \over 
              \left [ \xi  Q^2 +  k_{\perp}^2 / \bar \xi \right ] 
              \left [ \xi(x-\xi ) Q^2 +  x k_{\perp}^2 \right ]}
\right \} \, . \\[5mm]
           \label{26}   \end{equation}

The $k_{\perp}$-integral  diverges both in the  $k_{\perp} \to \infty$
and  $k_{\perp} \to 0$ limits.  The ultraviolet large-$k_{\perp}$
divergences (they are actually  irrelevant to  our analysis) 
 are removed by the $R$-operation,  while 
the low-$k_{\perp}$  collinear  divergences can be regulated by 
taking massive quarks.
In that  case, $k_{\perp}^2 \to k_{\perp}^2 + m^2$
and the small-$k_{\perp}$ divergence (collinear
 singularity)  is converted into
the  mass  logarithm $\ln(Q^2/m^2)$
generating the evolution of the pion distribution amplitude.
The  Sudakov effects  are
also related to  the  $1/k_{\perp}^2$ singularity.
It is easy to check that the coefficient in front
of $1/k_{\perp}^2$ in the singular part 
 \begin{equation}
M_a^{sing} (x,Q^2;\xi, k_{\perp}) = 
          -  {1\over x Q^2 }  
                 { Q^2  
                 \over k_{\perp}^2  \left [\xi  Q^2 + k_{\perp}^2 / \bar \xi\right ]
                           } \, 
\left ( {\bar \xi \over \bar x} \right ) \, \theta(\xi>x) 
\label{27} \end{equation}
has   the form of the 
product 
of the Born  term $1/\xi Q^2$
and the relevant  part 
\begin{equation}
V_a(\xi,x) = \left ( {\bar \xi \over \bar x}\cdot 
{ \, \theta(\xi > x) \over \xi - x} \right )_+
\label{28} \end{equation}
of the evolution  kernel (\ref{16}). 
Note,  that calculating 
the    evolution logarithm  $\ln Q^2/m^2$ 
from $d^2 k_{\perp}/k_{\perp}^2$, 
 one can  take    $k_{\perp}=0$ (``neglect $k_{\perp}$'')
 in all other places,
in particular, in the  denominator 
factor  $\xi  Q^2 + k_{\perp}^2 / \bar \xi$.
However,  nothing  prevents  us from going 
beyond the leading logarithm approximation.
Keeping   the $  k_{\perp}^2 $-terms,  
we can take into account those contributions which do not
have  logarithmic behavior  with respect to $m^2$ or $Q^2$.
We will see that among them, there are ``Sudakov''  terms with a 
specific  double-logarithmic
dependence on   the impact parameter $b_{\perp}$,
the variable which is 
 Fourier-conjugate to the transverse momentum $k_{\perp}$.
 To separate the  contributions related to 
 the  evolution kernel from those corresponding to
 Sudakov effects,
we first make the decomposition  
\begin{equation}
- {1\over   \left [\xi Q^2 + k_{\perp}^2/ \bar \xi \right ]  x Q^2 } 
=  \left ( {1 \over \xi Q^2 + k_{\perp}^2/ \bar \xi} - 
{ 1 \over xQ^2} \right )
{1 \over (\xi -x )  Q^2 + k_{\perp}^2/ \bar \xi}
\label{29} \end{equation}
and notice that    the denominator factor $\xi Q^2 + k_{\perp}^2/ \bar \xi$
reduces to $xQ^2$ when $\xi=x$ and $k_{\perp}=0$. 
Hence, we can  write
\begin{equation}
\begin{array}{rl} \displaystyle 
\frac{2 \pi}{Q^2} \, t_a^{sing}(x,Q^2) = - \int \limits_x^1 d\xi \int d^2 k_{\perp}  \ 
             {\bar \xi / \bar x \over  k_{\perp}^2 
\left [\xi Q^2 + k_{\perp}^2/ \bar \xi \right ] x Q^2} = 
  \int \limits_x^1 d\xi \displaystyle \int {d^2 k_{\perp}  \over \xi Q^2 + k_{\perp}^2/ \bar \xi } 
         \left \{  {\bar \xi / \bar x \over   k_{\perp}^2
               \left [ (\xi-x) Q^2 + k_{\perp}^2/ \bar \xi \right ] }
              \right . 
                       \\[5mm] \displaystyle
             \left .
     - \ \displaystyle \delta(\xi-x) \delta^2(k_{\perp})
          \int \limits_x^1 d\zeta \int d^2 \tilde k_{\perp}
             {\bar \zeta / \bar x  \over \tilde k_{\perp}^2
              \left [ (\zeta-x) Q^2 + \tilde k_{\perp}^2/ 
\bar \zeta \right] }\right \} \, . 
\end{array}
\label{30} \end{equation}
To disentangle the
product of the delta-functions in $\xi$ and 
$k_{\perp}$ variables, we rewrite  Eq.(\ref{30}) as 
\begin{equation}
\begin{array}{rl} \displaystyle
\int \limits_0^1 d\xi &\displaystyle \int {d^2 k_{\perp}  
\over \xi Q^2 + k_{\perp}^2/ \bar \xi } 
         \left \{ { 1 \over k_{\perp}^2}
             \left ( {(\bar \xi / \bar x) \, \theta(\xi > x) \over  
               (\xi-x) Q^2 + k_{\perp}^2/ \bar \xi }\right )_+
              \right . 
                       \\[5mm] \displaystyle
        &   \left .
        + \ \displaystyle \delta(\xi-x) \int \limits_x^1  \, 
              {\bar \zeta \over \bar x} 
              \left ( {1 \over k_{\perp}^2 \left [ (\zeta-x) Q^2 + 
k_{\perp}^2/ \bar \zeta \right ] }
            - \delta^2(k_{\perp})
          \int { d^2 \tilde k_{\perp}
             \over \tilde k_{\perp}^2
              \left [ (\zeta-x) Q^2 + \tilde k_{\perp}^2/ 
\bar \zeta \right] }\right )\,  d\zeta \,  
\right \}, 
\end{array}
\label{31} \end{equation}
where the combination  
\begin{equation}
\left ( {(\bar \xi / \bar x)  \, \theta(\xi > x)\over  
               (\xi-x) Q^2 + k_{\perp}^2/ \bar \xi  }\right )_+
\equiv 
{ (\bar \xi/ \bar x)  \, \theta(\xi > x) \over  
                (\xi-x) Q^2 + k_{\perp}^2/ \bar \xi  }
- \delta(\xi - x) \int \limits_0^1          
               { (\bar \zeta /\bar x)   \, \theta(\zeta > x)
               \over  (\zeta-x) Q^2 + k_{\perp}^2/ 
\bar \zeta  } \, d\zeta
\label{32} \end{equation}
is an analog    of the ``plus'' operation
for the case when the transverse momentum is  present.
Similarly,  the expression
\begin{equation}
{1 \over k_{\perp}^2 \left [ (\zeta-x) Q^2 + k_{\perp}^2/ \bar \zeta \right ] }
            - \delta^2(k_{\perp})
          \int { d^2 \tilde k_{\perp}
            \over \tilde k_{\perp}^2
              \left [ (\zeta-x) Q^2 + \tilde k_{\perp}^2/ \bar \zeta \right] }
\label{33} \end{equation}
can be interpreted as a   ``plus'' 
distribution with respect to  $k_{\perp}$.
Extracting the pure $1/ k_{\perp}^2$-singularity 
from the $(\cdots)_+$ term in Eq.(\ref{31}) 
\begin{equation}
{ 1 \over k_{\perp}^2}
             \left ( {(\bar \xi / \bar x) \, \theta(\xi > x) \over  
               \left [ (\xi-x) Q^2 + k_{\perp}^2/ \bar \xi \right ] }\right )_+
={1 \over {Q^2k_{\perp}^2}} 
             \left ( {(\bar \xi / \bar x) \, \theta(\xi > x) \over  
                (\xi-x)  }\right )_+  - 
            { 1 \over {Q^2} } \left ( {(\bar \xi / \bar x) \, \theta(\xi > x) \over  
             (\xi -x)  \left [ \bar \xi (\xi-x) Q^2 + 
k_{\perp}^2  \right ] }\right )_+  \,  , 
\label{34} \end{equation} 
we can write (\ref{31}) in the  impact parameter representation as
\begin{equation}
t_a^{sing}(x,Q^2) = \frac{1}{2 \pi}
\int \limits_0^1 d\xi \int  \ B(\xi; bQ) \
   \biggl [ V_a(\xi, x) L(bm) + E_a(x,\xi;bQ) + \delta (\xi -x) 
S_a(x, b Q) \biggr ]\,d^2 b_{\perp}  . 
\label{35} \end{equation}
The function  $B(\xi; bQ)$ gives  the  Born term in the $b$-space 
\begin{equation}
B(\xi; bQ) =
   \frac{1}{2 \pi}\int  { e^{- i k_{\perp} b_{\perp} }
 \over \xi Q^2 + k_{\perp}^2 / \bar \xi} \,  d^2 k_{\perp}
   =  \bar \xi  K_0 \left ( b Q \sqrt{\xi \bar \xi} \right ),
\label{36} \end{equation}
where $b= |b_{\perp}|$ and  $K_0(z)$ is the modified Bessel function.
By   $L(bm)$  we denote a  regularized version of the integral
resulting from the first term in 
Eq.(\ref{34}):  
\begin{equation}
L(bm) = {\rm Reg}_{(m)} \left \{ \frac{1}{2 \pi} \int d^2 k_{\perp} 
{ e^{i k_{\perp} b_{\perp} } \over k_{\perp}^2} \right \} .
\label{37} \end{equation} 
In particular, if the integral  is regulated by $1/k_{\perp}^2 \to
1/(k_{\perp}^2+m^2)$, then $L(bm)=K_0(bm)$. 
The  function $L(bm)$  
 contains the mass logarithm  $\ln(mb)$ multiplied by 
the  relevant  part $V_a (\xi,x)$ of the evolution kernel.
 As discussed in the preceding section,
the mass singularity $\ln(m)$ must   be absorbed 
(in the form  $\ln(m/\mu)$, where $\mu$ is 
the  factorization scale)
into the 
redefinition of the distribution amplitude: 
$\varphi_{\pi}(x) \to \varphi_{\pi}( x; \mu)$.
The second term in  Eq.(\ref{34}) is 
given by the  function   $E(x,\xi; bQ)$  
which also contains the evolution kernel $V_a (\xi,x)$ 
\begin{equation}
\begin{array}{rl}
E_a(x,\xi; bQ)=& \displaystyle
 - \frac{1}{2 \pi} \int   e^{i k_{\perp} b_{\perp}} 
             \left ( {(\bar \xi/ \bar x)
 \, \theta(\xi > x) \over 
              ( \xi-x )[\bar \xi (\xi-x) Q^2 +
 k_{\perp}^2 ] }\right )_+  d^2 k_{\perp}
              =  -  \left [ {\bar \xi \over \bar x} \ {
               \, \theta(\xi > x) \over \xi-x} \ 
                    K_0 \left ( b Q  \sqrt{(\xi-x) \bar \xi} \right )
 \right ]_+
                  \, . 
\end{array}
\label{38} \end{equation}
It is easy to notice that both the Born term
$B(\xi;bQ)$ and the evolution-related terms $L(bm)$ and 
$E_a(x, \xi; bQ)$
exponentially decrease  at large $b$,
since the function 
$K_0 \left ( b \ldots  \right ) $ behaves like  
$ \exp(- b \ldots  ) $  in this limit. 
 On the other hand, the  ``Sudakov'' term 
\begin{equation}
S_a(x;bQ) = 
         \frac{1}{2 \pi} \int d^2 k_{\perp} { e^{i k_{\perp} b_{\perp}} - 1 \over k_{\perp}^2}
             \int \limits_x^1 \left ( {\bar \zeta^2 \over \bar x} \right ) \ 
                 {d \zeta \over \bar \zeta (\zeta-x) + k_{\perp}^2 / Q^2 }   
\label{39} \end{equation}
accompanied by 
$\delta(\xi-x)$  in Eq.(\ref{35})
has a completely different behavior at large $b$.
Indeed, changing the variable $\zeta$ in the 
above integral as $1-\zeta=y \bar x$,
we rewrite Eq.(\ref{39}) in the form
\begin{equation}
S_a(x;Qb) =   \frac{1}{2 \pi}
\int d^2 k_{\perp} { e^{i k_{\perp} b_{\perp}} - 1 \over k_{\perp}^2}
             \int \limits_0^1 
       { { y^2 d y} \over y \bar y + k_{\perp}^2 / {\bar x}^2 Q^2}
      \,  \equiv   s ( \bar x Q b ).
\label{40} \end{equation}
According to this  representation, 
 the function $s(\bar x Q b)$ vanishes as  $b\to 0$. 
In the opposite limit of large impact parameters,
it has a  double-logarithmic dependence on $b$.
To see this, we integrate first over $y$ and then over $k_{\perp}$
taking into account
that the factor  $(e^{i k_{\perp} b_{\perp}} - 1)$ provides,
 in the limit of large $b$, an effective IR cut-off 
at $k_{\perp} \sim 1/b$. As a result,   
we obtain the large-$b$ behavior of $s ( \bar x Q b ) $
\cite{bottssterman} 
\begin{equation}
s ( \bar x  Q b )  \approx  \frac{1}{2 \pi} 
\int d^2 k_{\perp} { e^{i k_{\perp} b_{\perp}} - 1 \over k_{\perp}^2}
                      \,  \mbox{ln} \left ({ \bar x Q \over k_{\perp} } \right )
           \approx    \int \limits_{1/b}^{} {d k_{\perp} \over k_{\perp}} \, 
                       \mbox{ln} \left ({ k_{\perp} \over \bar x Q} \right )
            \approx   -  \frac1{2} \; \mbox{ln}^2 (\bar x  Q  b), \ \ \ 
1/\Lambda_{QCD} \gg b \gg 1/Q.
\label{41} \end{equation}
To be on safe side, we included the $1/\Lambda_{QCD} \gg b$  restriction 
to emphasize that these  results are only valid in the  region
where one can trust  pQCD expressions for quark and gluon 
propagators.  

Integrating $s ( \bar x  Q b )$ with the Born term
gives, for small $x$,  a  negative double logarithm $-\frac1{2} \ln^2 x$.
As discussed above,  such a correction is expected when one
uses $xQ^2$ as the factorization scale.  
Indeed, for small $x$, the Born term is a function 
of $x b^2 Q^2$. Hence, the choice $\mu^2 =1/b^2$ 
is essentially equivalent to setting $\mu^2 \sim  x Q^2$.

In ref.\cite{bottssterman}, it was shown that
the $b$-space double logarithms  exponentiate
in higher orders.
In the double logarithmic approximation,
they give the  suppression factor 
\begin{equation}
 \exp \left \{ - \frac{\alpha_s}{ 4\pi} \, 
C_F \ln^2 (\bar x  Q b)\right \} 
\label{42} \end{equation}
for large $b$.
 The running of the coupling constant induces the next-to-leading 
logarithms 
(cf. \cite{poggio,smilga}). To get them, one should 
put  $\alpha_s(k_{\perp}^2) = 4\pi/(\beta_0 
\ln k_{\perp}^2/\Lambda^2)$ under the integral:
\begin{equation}
\alpha_s \,C_F \,  s ( \bar x  Q b )  \to \frac{C_F}{2 \pi } \int d^2 k_{\perp} 
{ e^{i k_{\perp} b_{\perp}} - 1  \over k_{\perp}^2}
\,  \alpha_s(k_{\perp}^2)
            \int \limits_0^1  \, 
       { { y}^2 \, dy \over y \bar y + k_{\perp}^2 /
 {\bar x}^2 Q^2} \,          .
\label{43} \end{equation}
In general, the Sudakov effects are governed by the  $eikonal$ 
\cite{catren,collins,bottssterman}
(or $cusp$ \cite{korsud,korradW,korrad92} )  
anomalous dimension 
\begin{equation}
\Gamma_{cusp} (\alpha_s) = \frac{C_F \alpha_s}{ \pi} \left \{
1  + \frac {\alpha_s}{\pi} \left [  N_c \left (\frac{67}{36}
 - \frac{\pi^2}{12} \right ) -\frac5{18} N_f\right ] +\ldots \right \} \,  .
\label{44} \end{equation} 
Clearly, only the  $\alpha_s$ term  of $\Gamma_{cusp} (\alpha_s)$
manifests itself in a one-loop calculation. 
To get further corrections \cite{bottssterman},  one should  substitute
$C_F \alpha_s/ \pi$ in Eq.(\ref{43})
by  $\Gamma_{cusp} (\alpha_s) $
 and also use a two-loop expression for $\alpha_s(k_{\perp}^2)$
and $s(\bar x Qb)$  \cite{bottssterman}.  
Here, we restricted our analysis by the one-loop level.

\subsection{ Vertex correction for the real  photon}

For the real photon, 
 the contribution of the  vertex correction diagram $3b$ is given by
\begin{equation}
M_b (x,Q,  \xi,k_{\perp})
     =  \displaystyle {1 \over x Q^2 } \left \{ {\xi \over x} \cdot
                  { (x-\xi)Q^2 + x k_{\perp}^2 \over k_{\perp}^2
                     \left [ \xi(x-\xi )Q^2 +  
x k_{\perp}^2 \right ] }
                                            \right \} \theta (\xi<x) \ . 
\label{45} \end{equation}
Again, we concentrate on the term  singular at $k_{\perp} =0$.
It is convenient to split it  into two parts.
The first part is obtained by taking $xQ^2$ 
from the $(x-\xi)Q^2$ term in the numerator
and the second one by taking $(-\xi Q^2)$.
We  represent the first part  as 
\begin{equation}
\begin{array}{rl}
&\displaystyle  \left (  {\xi \over x} \right ) \cdot 
                        {1 \over {k_{\perp}^2}\left [ \xi(x-\xi ) Q^2 + 
 x k_{\perp}^2 \right ] }   = \\[0.5cm]
&\displaystyle {1 \over k_{\perp}^2} \cdot \left ( 
{1 \over \xi Q^2+k_{\perp}^2/\bar \xi } \right  ) \cdot  
 {\xi^2 / x  \over \xi(x-\xi ) +  x k_{\perp}^2 /Q^2 } 
 +  { {\xi / x}\over [\xi \bar  \xi Q^2+k_{\perp}^2 ] 
   [ \xi(x-\xi )Q^2  +  x k_{\perp}^2] } \,  . 
\label{46}
\end{array}
\end{equation}
The last  term here  produces no divergences both   for large and small $k_{\perp}$.
The $1/k_{\perp}^2$ singularity is contained in the 
first term which 
we arranged  to have  a form of a product  
of the same  Born term 
$1 / (\xi Q^2+k_{\perp}^2/\bar \xi)$  with 
a factor looking like a $k_{\perp}$-modified evolution kernel.
Then we  write this factor   as a sum of a 
``plus'' term and a $\delta(x-\xi)$ term:
\begin{equation}
 {(\xi^2 / x)  \, \theta( \xi <x) 
   \over  \xi(x-\xi )  +  x k_{\perp}^2/Q^2   } =
\left  ( {{(\xi^2 /x)\, \theta( \xi <x)  } \over 
      {\xi(x-\xi )  +  x k_{\perp}^2/Q^2   }} \right  )_+ 
   + \delta(x-\xi) \int\limits_0^1  \, {(\zeta^2 / x ) \,  
                    \theta( \zeta <x) 
       \over  \zeta(x-\zeta )  +  x k_{\perp}^2/Q^2 } \, 
d\zeta \, . \label{47}
\end{equation} 
As a result, the total contribution associated 
with the  $k_{\perp} =0$ singularity 
 can be written as
\begin{equation}
\begin{array}{rl} &
 t_b^{sing}(x,Q^2) =  \displaystyle \int \limits_0^1 
d\xi \displaystyle \int {d^2 k_{\perp}  
\over \xi Q^2 + k_{\perp}^2/ \bar \xi } 
         \left \{ { 1 \over k_{\perp}^2}
             \left ( {( \xi^2 /  x) \, \theta(\xi < x) \over  
               \xi (x- \xi)  + x k_{\perp}^2/Q^2   }\right )_+
              \right . 
                       \\[5mm] \displaystyle
        & \hspace{5mm}  \left .
        + \ \displaystyle \delta(\xi-x) \int \limits_0^x  \, 
              { \zeta^2 \over  x} 
              \left ( {1 \over k_{\perp}^2 \left [ \zeta (x - \zeta)  + 
x k_{\perp}^2/Q^2 \right ] }
            - \delta^2(k_{\perp})
          \int { d^2 \tilde k_{\perp}
              \over \tilde k_{\perp}^2
              \left [ \zeta (x - \zeta)  + x \tilde 
k_{\perp}^2/Q^2 \right] }\right ) \, d\zeta  
\, \right \}, 
\end{array}
\label{48}
\end{equation}
where the $\delta^2(k_{\perp})$ term 
comes from  the second, ``$-\xi Q^2$''  
part of the original expression (\ref{45}).

From this decomposition, we obtain the mass singularity 
term 
\begin{equation}
\left ( {\xi \over x} \, {\theta(\xi <x) 
\over  (x-\xi) }\right )_+ L(bm) \equiv  V_b(\xi, x)\,  L(bm)\,  ,  
\label{49} \end{equation}   
the evolution-related  contribution 
\begin{equation}
E_b (x,\xi;b) =   
    - \left [ {\xi \over x} \, {\theta(\xi <x) \over  (x-\xi) }
 K_0(bQ\sqrt{ \xi (x-\xi)/x} \, ) \right ]_+       
\label{50} \end{equation}
and the Sudakov term
\begin{equation}
S_b(x;bQ) = 
         \frac{1}{2 \pi} \int d^2 k_{\perp} { e^{i k_{\perp} b_{\perp}} - 1 \over k_{\perp}^2}
             \int \limits_0^x \left ( {\zeta^2 \over  x} \right ) \ 
                 {d \zeta \over  \zeta (x- \zeta) 
+ x k_{\perp}^2 / Q^2 } = s(\sqrt{x} Qb)\,  .
\label{51} \end{equation}
For large $b$, the latter behaves like 
\begin{equation}
S_b (x;bQ) \approx  -  \frac1{2} \ln^2 \left (\sqrt{x} Qb  \right ).
\label{52} \end{equation}
By analogy with $S_a (x;bQ)$ which is a function of $\bar  x Qb$
we might expect that $S_b (x;bQ)$ should be  a function of $xQb$.
Our calculation above shows that $S_b (x;bQ)$ is a function 
of $\sqrt{x} Qb$. 
That this result is not unreasonable, can be  justified in the following way.
Note,  that for small $x$, both the Born term $B(x;bQ)$ 
and  our $S_b(x;bQ)$ are the functions of the 
same combination $xb^2 Q^2$. Hence, integrating the product 
$B(x;bQ) S_b(x;bQ)$ over $b$ 
just gives $1/Q^2$ multiplied by a constant factor:
no $\ln x$ terms are produced.  On the other hand, a $\ln ^2 x$ term would appear 
if   $S_b(x;bQ)$ would behave  like  $\ln ^2 (xQb)$ for large $b$.  
The  explicit expression for 
diagram $3b$ given in ref.\cite{braaten} 
has no $\ln^2 x$ terms.

\subsection{Box  and self-energy diagrams}

In Feynman gauge, the box diagram $3c$  contribution in QCD 
\begin{equation} 
M_c (x,Q ; \xi, k_{\perp})= 
       {1\over xQ^2} \left \{
        { x\left ({\bar \xi}^2 Q^2 + k_{\perp}^2 \right ) \over 
       \bar x k_{\perp}^2 \left [ \xi \bar \xi Q^2 + k_{\perp}^2 \right ]}
        - { (x - \xi)^2 Q^2  + x k_{\perp}^2 \over 
     \bar x k_{\perp}^2 \left [ \xi (x-\xi) Q^2 + x k_{\perp}^2 \right ]}
                                    \, \theta(\xi<x)     \right \}
  \label{53} \end{equation}
only by a numerical  factor differs from that in a model
with scalar or pseudoscalar gluons, in which 
Sudakov effects are absent.
Hence,  the $k_{\perp}=0$ singularity  produces only  the evolution effects:
\begin{equation} 
M_c(x;\xi, k_{\perp}) = \frac1{k_{\perp}^2} \,  V_c(x,\xi) \, \frac1{\xi Q^2} + 
\ldots 
  \label{54} \end{equation}
where $V_c(x,\xi)$  is the relevant part
\begin{equation} 
V_c(x,\xi) = \frac{\xi}{x} \, \theta( \xi < x) + \frac{\bar \xi}{\bar x} \, 
\theta( \xi > x) 
  \label{55} \end{equation}
 of  the evolution kernel. 
Note, that  $V_c(x,\xi)$ does not have a ``plus'' form by itself.
The missing $\delta(x-\xi)$ terms are provided by two 
quark self-energy diagrams  $3d,e$
\begin{equation}
\begin{array}{rl}
M_{d+e} =& - \displaystyle
           {1 \over xQ^2} \,  \delta (x-\xi) \,  {1 \over k_{\perp}^2} 
                \int_0^1  \left [ {{\bar \zeta}\over{\bar x}} \,
\theta(\zeta>x)
               + { \zeta\over x} \, \theta(\zeta<x) \right ] \, d\zeta
               = - {1 \over 2xQ^2  k_{\perp}^2} \, \delta(x-\xi) . 
\end{array}
\label{56} \end{equation}
The third self-energy diagram $3f$ has only the  UV divergence:
\begin{equation}
M_f     =
           -{1\over x Q^2} \, \delta(x-\xi)
              \int_0^x 
              {\zeta/x \over  \zeta (x-\zeta) Q^2 + x k_{\perp}^2} 
\,d\zeta \, d^2 k_{\perp}.
\label{57} \end{equation}
Combining  evolution kernels from all the  diagrams above,
one obtains  the total evolution kernel $V(\xi,x)$ (\ref{16}).

\subsection{Standard $vs.$ modified factorization}

Summarizing the findings of the
previous subsections, we  write the sum of
the lowest-order term and  one-loop 
diagrams  in the  impact 
parameter representation as
 \begin{eqnarray}
&& \lefteqn{ F_{\gamma^* \gamma \pi^0}(Q^2) =
\frac{4\pi}{3}\int \limits_0^1  \, 
\Biggl  \{\frac1{xQ^2}  + \frac{\alpha_s}{2 \pi} \,  C_F  
\int \limits_0^1 d\xi \int 
 \ B(\xi; bQ) \
 \biggl [  V(\xi, x)\ L(bm) + E(\xi, x ;bQ) 
 \label{35B} } \\   && \hspace{6cm} +  
 \delta (\xi -x) 
S(x, b Q) + R(\xi, x;bQ)\biggr ] 
 \,\frac{d^2 b_{\perp}}{2 \pi}
  \, \Biggr \} \, \varphi_{\pi}(x) \, dx \, ,  
\nonumber \end{eqnarray}
where $ B(\xi; bQ)$ is the $b$-version of 
the Born term (\ref{36}),
$V(\xi, x)$ is  the total evolution kernel,
$E(x,\xi;bQ)$ is the sum of the evolution-related terms
like (\ref{38}), (\ref{50}), 
$S(x, b Q) $
is the total Sudakov term given by Eqs.(\ref{40}), (\ref{51})
and $R(\xi, x;bQ)$ accumulates all the remaining
contributions coming from terms regular at $k_{\perp}=0$. 
Integrating over $b$ and specifying the prescription for the
renormalized distribution amplitude $\varphi_{\pi}(x;\mu)$,
one would get the result (\ref{14}) of the 
standard factorization scheme. 
In particular, the term $\frac12 \ln^2 x $, 
most sensitive to the width
of the distribution amplitude $\varphi_{\pi}(x;\mu)$,
comes  from a negative contribution 
$-\frac12 \ln^2 x $
due to the Sudakov term $S(x, b Q)$ and a 
positive contribution  
$ \ln^2 x $  coming from the $m$-independent part of
the convolution 
\begin{equation}
\int_0^1 d \xi \, \int 
B(\xi; bQ) \otimes V(\xi,x)  L(bm) \,
 \frac{d^2 b_{\perp}}{2 \pi}  = \frac1{xQ^2}
\biggl  \{ \left (\frac32 + \ln x  \right )
\ln (Q^2/m^2) +\ln^2 x + f(x)  \biggr \} .
\label{mind} \end{equation}
This convolution    contains also  terms denoted by $f(x)$
which are less singular at $x=0$.
The total sum vanishes when integrated 
with the non-evolving asymptotic distribution amplitude
$\varphi_{\pi}(x)$. 
It does not vanish, however, when integrated 
with DA's differing from $\varphi_{\pi}^{as}(x)$.

The  logarithmic mass singularity $\ln m$ 
contained in the evolution term $V(\xi, x)\, L(bm)$ 
is eliminated by absorbing it into the renormalized
DA. 
The procedure used 
in the modified factorization approach of 
refs.\cite{bottssterman,listerman} is to
  absorb  $\ln(mb)$.  As a result, one obtains
 the pion distribution
amplitude $\varphi_{\pi}( x; 1/b)$ normalized at the scale 
$\mu = 1/b$. Making such a choice, 
 one should realize  that
$b$ is  an integration variable and, 
to preserve the acquired precision,
one must use the evolution equation  to get 
 $\varphi_{\pi}( x; 1/ b)$ for all relevant values of $b$.
In particular, if   the  
distribution amplitude is  
assumed to have a CZ-type shape  
for large $b$,  it 
 should be evolved 
towards the asymptotic 
shape  for smaller $b$ using  Eq.(\ref{21}).
Modelling  $\varphi_{\pi}( x; 1/ b)$ by 
a function of $x$ only amounts to neglecting
the $m$-independent part of the convolution 
$B(\xi; bQ) \otimes  V(\xi, x)L(bm)$  (\ref{mind}).
 As noted before, this contribution
contains $ \ln^2 x $, hence,   for extremely
wide distribution amplitudes 
it can exceed that coming from the Sudakov 
term which only contains $ (- \frac12 \ln^2 x) $.

In the formal 
 $b=0$ limit, the function  $\varphi_{\pi} ( x; 1/b)$  
evolved according to the
leading logarithm approximation formula (\ref{21}), coincides
 with $\varphi_{\pi}^{as}( x)$.  
However,   
the  function   $E(x,\xi; bQ)$  
also develops a  logarithmic 
singularity for small $b$,  because 
$$K_0 \left ( Q b \ldots  \right ) =
- \ln (Qb) +\ldots $$
for small $b$. 
Hence,    two $\ln(b)$ singularities present in Eqs.(\ref{35}),(\ref{35B})
 compensate each other in the $b \to 0$ limit  and the net 
coefficient in front of the evolution kernel
is $\ln(Q/m)$:  
the distribution amplitude evolves in fact only to
the scale $b_{min} \sim 1/Q$ corresponding to 
the  resolving power of the external probe.
Absorbing  $\ln(Q/m)$ into the 
renormalized distribution amplitude
one  would  get $\varphi(x) \to \varphi( x; Q)$,
with  the large external  momentum $Q$
serving now as a factorization scale.
Such a choice 
is usually made in the standard 
factorization approach, in which  
 $\mu$ is either
a fixed constant, $e.g.,$ $\mu= 1\, GeV$
or  proportional to the  external momentum,
 $\mu=aQ$,  with $a$ being a fixed number. 
In particular,  one can optimize the choice of 
 the parameter $a$   by taking the value producing 
the shape of $\varphi(x;1/b)$ 
averaged over the essential region 
of the $b$-integration.
Another point is that the pQCD evolution of $\varphi(x;\mu)$
is reliable only in a restricted region $\mu \gtrsim \mu_0$.
Since  the modified factorization involves integration 
over all $b$, we  formally 
need to  know   the distribution amplitude $\varphi(x;1/b)$ 
outside the perturbative region $ b \lesssim 1/ \mu_0$.
One should remember, however, that the Born term $K_0(Qb \sqrt{x \bar x})$
for finite $x$ 
  exponentially suppresses  the large-$b$ region.
As a result,  essential  impact parameters $b$ are $\sim 1/Q$.
The   suppression by the Born term  disappears  
 for small $x$ when   
 the effective scale becomes $1/\sqrt{x  Q^2}$ rather than 
$1/Q$. In this case, the suppression of the large-$b$ region
is provided by the exponentiation  of the
Sudakov terms which is 
 the crucial element of the modified 
factorization approach \cite{bottssterman,listerman}.
 As a result of the exponentiation, the series of $[\alpha_s
\ln^2 (Qb)]^n$ terms, each of which tends
to infinity as $b \to \infty$, is substituted 
by the exponential of Eq.(\ref{42}) type 
rapidly vanishing with growing $b$.
Of course, for finite $x$, the Born term $K_0(Qb \sqrt{x \bar x})$
 provides  even stronger 
  suppression of the large-$b$ region 
and the influence of the Sudakov factor  is minor. 
 Only for small $x$   Sudakov effects become 
important.  The relevant combination $\bar x  Qb$ 
in the Sudakov term  of the diagram $3a$ 
converts into $Qb$,  and the exponentiated Sudakov factor 
plays a  primary role in  squeezing  the size of essential 
impact parameters. 
A special role of the small 
$x$-valus  in the $b_{\perp}$-integration
is reflected by the $- \frac12 \ln^2 x$ term 
resulting from the convolution of the Born 
term with the one-loop Sudakov factor:
\begin{equation}
\frac1{2 \pi} \int B(x; Qb)\,  S(x ; Qb) \, 
d^2b_{\perp} =  \frac1{xQ^2} \left ( - \frac12 \ln^2 x - g(x) \,\right ) \,  ,
\label{BS}
\end{equation}
where $g(x)$ stands for less singular terms.
After integration with the asymptotic  
distribution  amplitude,  
the $ (- \frac12 \ln^2 x - g(x))$ term gives approximately
 $-9/4 + 0.05$,
to be compared with the magnitude $-5/2$ of the total
one-loop correction (see discussion
after Eq.(\ref{19})). Hence,  the  
total one-loop correction 
in the case of the asymptotic DA
is very close 
to the contribution of the Sudakov term alone
(the deviation is only  12\%).
If the  higher-loop corrections can be also approximated
by the Sudakov contribution, then the 
exponentiated form would produce  
the  all-order result in a rather compact form. 

Discussing the numerical significance of the
Sudakov terms,  we should keep in mind that all
the logarithmic enhancements 
$\ln^2 (Qb)$ are perfectly integrable and that
the region 
of small $x$, where the Sudakov terms are 
important, is  small itself:  after  $b_{\perp}$- and $x$-integrations, 
 there are no  especially large contributions in the final result.
The total one-loop correction
is only about $20 \%$. Hence, the exponentiation
of the Sudakov terms 
would alter the one-loop corrected 
result for the form factor 
 by just a few percent, 
which is similar to the 
accuracy  of approximating the total 
contribution 
by the Sudakov term at one loop.
Note also that  a few percent change 
may be smaller than the contribution generated by 
the one-loop terms $E(x,\xi, Qb)$,  $R(x,\xi, Qb)$
and the effects due to the $b$-dependence
of the renormalized distribution amplitude
$\varphi_{\pi}(x;1/b)$. Moreover, for a wide DA, the
latter are comparable to 
or exceeding the Sudakov contributions.
In principle, one can try to explicitly 
 include these corrections 
within the MFA framework, but the result would not
have a simple form anymore. In this situation,
instead of dealing with convolutions of   Bessel functions, 
one may prefer to use the result  (\ref{14})
of the standard factorization approach
which has a simple form  with easily controllable
accuracy.
Another bonus of using the SFA is the ability of 
$\varphi_{\pi}(x;Q)$
to fully absorb the  necessary nonperturbative information:
increasing $Q$ we do not need to make any assumptions
about the shape of $\varphi_{\pi}(x;\mu)$ at smaller values 
$\mu <Q$ of the factorization scale $\mu$.

\section{Inclusion of primordial transverse momentum}

\subsection{ Brodsky-Lepage interpolation}

Despite  our persistent efforts,  we  failed  so far 
to find any traces of contributions
capable of producing  a series of transverse-momentum-related
power corrections to the leading
pQCD result.  Recall that we investigated  
first  the  higher-twist  contributions
 due   to  operators with  contracted 
covariant derivatives
$D^{\mu} \ldots  D_{\mu}$
which are the standard candidates to describe the 
$k_{\perp}$-effects in the OPE-like factorization 
approaches. 
We observed that, 
  for the simplest handbag diagram,
these operators do not produce the expected 
infinite  chain  of $(1/Q^2)^n$ power  
corrections.  Then we studied
one-loop  
 radiative corrections  in the Sudakov 
and impact-parameter representations. Our results   are 
in full accord with the corresponding expressions  of the MFA
\cite{bottssterman,listerman}. But they  also completely 
agree with the one-loop results \cite{AuCh81,braaten,kmr} 
of the SFA, $i.e.,$ 
they  do not  contain any   power  corrections.
Nevertheless,   $F_{\gamma^*\gamma \pi^0}(Q^2) \sim 1/Q^2$   cannot 
be a  true behavior 
of $F_{\gamma^*\gamma  \pi^0 }(Q^2)$  in the low-$Q^2$
region,  especially since   the $Q^2=0$ limit of 
$F_{\gamma^*\gamma  \pi^0 }(Q^2)$ 
is known to be finite and 
 normalized by the $\pi^0 \to \gamma \gamma$ decay rate. 
The  value of $F_{\gamma^*\gamma  \pi^0 }(0)$ in QCD \cite{f0}
is fixed by the axial  anomaly \cite{anomaly}
\begin{equation}
F_{\gamma^*\gamma  \pi^0 }(0) = \frac1{\pi f_{\pi}}.
  \label{58} \end{equation}
If the shape of the pion DA is specified, 
the large-$Q^2$ behavior is also known.
For the asymptotic DA,
\begin{equation}
F_{\gamma^*\gamma  \pi^0 }^{as}(Q^2) = \frac{4\pi f_{\pi}}{Q^2}.
  \label{59} \end{equation}
Long ago, Brodsky and Lepage \cite{blin}  
 proposed the   interpolation formula
\begin{equation}
F_{\gamma^*\gamma  \pi^0 }^{int,BL}(Q^2) = 
{ {1} \over {\pi f_{\pi} 
\left (1+{{Q^2}\over{4 \pi^2 f_{\pi}^2}} \right )}} \equiv 
{{1} \over {\pi f_{\pi}(1+Q^2/s_0)} },
 \label{60} \end{equation}
which reproduces both the $Q^2 =0 $ value (\ref{58})
and the high-$Q^2$ behaviour  given by Eq. (\ref{59}).
The BL-interpolation formula (\ref{60}) has a  monopole form 
 with 
the scale $s_0 = 4 \pi^2 f_{\pi}^2 \approx 0.67 \, GeV^2$ 
 numerically   close to the $\rho$-meson mass squared:
$m_{\rho}^2 \approx 0.6 \, GeV^2$.
Thus, the BL-interpolation suggests a form similar to that based on  
the VMD expectation 
$F_{\gamma^*\gamma  \pi^0 }(Q^2) = 1/[\pi f_\pi (1+Q^2/m_{\rho}^2)]$.
In the VMD-approach,   the $\rho$-meson mass $m_{\rho}$ 
serves as a parameter  determining  the pion charge radius,
and it is only natural to 
expect that the tower of $(s_0/Q^2)^N$-corrections
suggested by the BL-interpolation formula can be 
explained by 
 intrinsic transverse momentum effects.
The only  problem  is   {\it how} to get Eq.(\ref{60}) 
(or anything similar to it) 
from QCD, $i.e.,$ how 
to construct an  expression which would 
provide a good model both in 
perturbative and nonperturbative regimes. 
Before proposing our variant of the solution to this problem, 
let us discuss briefly  two recent  attempts \cite{kroll,huang} to 
include  intrinsic transverse momentum
effects into the description of the  $\gamma^*\gamma  \pi^0$
form factor.

\subsection{Extrapolation of perturbative results}

As emphasized above, despite the fact that 
the denominator 
 of the Born term $1/(
\xi Q^2 + k_{\perp}^2 / \bar \xi )$ is  
$k_{\perp}$-modified 
compared to its collinear 
approximation $\xi Q^2$,  convoluting $B(\xi ; bQ)$ with 
$S( \xi; bQ)$ one would enjoy  no power modifications of the
canonical $1/Q^2$-behavior, $i.e.,$ the transverse-momentum 
effects included in the Sudakov term and 
other one-loop corrections do not 
correspond to any higher-twist contributions.
The obvious reason is that, apart from the 
IR regulator  mass $m$ (producing  a 
 logarithmic dependence $\ln m$ which is 
absorbed into $\varphi (x;\mu)$), 
 the large momentum $Q$ is the only 
scale that appears in the relevant 
$k_{\perp}$-integrals.  

In general,  the fact that some contribution
is written  as an integral over  the 
 transverse momentum  $k_{\perp}$ 
or the impact parameter $b_{\perp}$  
  does not  necessarily mean  that 
something  beyond the leading twist is 
included. To illustrate this point, we note that 
 even the 
lowest-order, ``purely collinear''
contribution (\ref{8}) 
 can be written  in the  
 impact-parameter  representation.
A possible form is suggested by the one-loop 
calculation:
\begin{equation}  
F_0 (Q^2) = \frac{2}{ 3} \int_0^1 dx  \int 
\bar x  K_0 \left (  \sqrt{x\bar x b^2 Q^2 } \right ) 
\varphi_{\pi} (x) \, d^2 b \,  , 
 \label{61} \end{equation}
where  $\bar x K_0(\sqrt{x \bar x b^2 Q^2 })$  is 
 the  impact-parameter profile
of the  modified
propagator $1/(x Q^2 + k_{\perp}^2/\bar x)$ (see Eq.(\ref{36})). 
Though the $b$-version of the quark
 propagator explicitly depends on $b$, 
integrating over $b$  in Eq.(\ref{61}) gives 
 a simple power result $1/Q^2$   without 
any subleading  power corrections.
This phenomenon can be traced to 
the absence of the $b$-dependence in 
the distribution amplitude.
In the momentum representation,  
Eq.(\ref{61})  is equivalent to using 
$\varphi_{\pi} (x) \delta^2(k_{\perp})$ 
for the $\bar q q \pi$ vertex:
\begin{equation}
F_0 (Q^2) = \frac{4 \pi}{3} 
\int_0^1 dx \, \int  {{\varphi_{\pi} (x) \delta^2(k_{\perp})  }
\over {x Q^2 + k_{\perp}^2/\bar x  }} \, 
 d^2 k_{\perp} \,  . 
\label{62} \end{equation} 

However, as we have seen  in the preceding section,  
radiative corrections
generate  terms
with less trivial  $k_{\perp}$-dependence.
In particular, the 
one-loop correction contains  
$\alpha_s /k_{\perp}^2$  terms.
As a result, the $k_{\perp}$-dependence of the  $\bar q q \pi$ vertex
at one loop is 
\begin{equation} 
\varphi_{\pi} (\xi) \delta^2(k_{\perp}) + \frac{\alpha_s}{(2 \pi)^2 
k_{\perp}^2} \int_0^1 V(\xi, x) \,  \varphi_{\pi} (x) \, dx  + \ldots \,  . 
\label{63} \end{equation} 
In the impact parameter representation, 
the sum of $\delta^2(k_{\perp})$ and $1/{k_{\perp}^2}$
terms  is converted into a more  suggestive  combination 
\begin{equation}
 \varphi_{\pi} (\xi)  -  
 \frac{\alpha_s}{2 \pi}\ln(bm)
 \int_0^1 V(\xi, x) \, \varphi_{\pi} (x) \, dx , 
\label{64} \end{equation} 
which  can be understood   as the two first terms 
of the $\alpha_s$-expansion of the  expression
for  the leading-logarithm evolved distribution amplitude
$\varphi(\xi, 1/b)$  written symbolically as 
$$
\exp \left [ - \frac{\alpha_s}{2 \pi}\ln(bm) V \right ] \otimes \varphi \, .
$$

Since all  the  conclusions made from the studies of 
one-loop corrections  are based on perturbative analysis,
strictly speaking,  they   are  only applicable  to 
transverse momenta which are large enough
\footnote{In particular, speaking about the 
 double-logarithmic behavior  
``at large $b$''  we imply  that $b$ may be 
much larger numerically 
than $1/Q$ but  is  still within the pQCD 
applicability range.}.
Furthermore, there are no special reasons to 
expect that   formulas derived for 
 momenta  $k_{\perp}$ generated by
perturbative gluon radiation  
 are still true 
in the  small-$k_{\perp}$ 
region dominated by 
 primordial (or intrinsic) transverse momentum. 
Still,  it is tempting to 
extend  the  leading-logarithm convolution formula
\begin{equation}
F (Q^2) = \frac2{3}  \int_0^1 dx   \int K_0(\sqrt{x \bar x } Qb) \, 
\varphi (x; 1/b) \, d^2 b
\label{65} \end{equation}
 into the nonperturbative region. To do this, we should 
substitute the distribution
amplitude $\varphi (x; 1/b)$ by a function 
which  reflects (or models) the  nonperturbative $b$-dependence.
   
In the  light-cone approach 
 \cite{bl80},  the basic object is  
 the wave function $\Psi(x, k_{\perp})$ which depends both on
the fraction variable $x$ and transverse 
momentum $k_{\perp}$.  In QCD,  it is customary 
 to  split  $\Psi(x, k_{\perp})$  into  two 
components. The soft component $\Psi^{soft}(x, k_{\perp})$
is due to the nonperturbative
part of the QCD interaction and  its   width  is determined
by the size of the relevant $\bar q q$
bound state. It is  expected that 
$\Psi^{soft} (x, k_{\perp})$ rapidly ($e.g.,$ exponentially)
 decreases for large  $k_{\perp}^2$. 
In our perturbative lowest-twist treatment   above, 
the soft wave function $\Psi^{soft}(x, k_{\perp})$ 
was imitated  by $\varphi_{\pi}(x) \delta^2 (k_{\perp})$.
The pQCD interaction (gluon radiative corrections) 
produces the hard 
component $\Psi^{hard} (x, k_{\perp})$ which behaves like
$\alpha_s/k_{\perp}^2$ at large $k_{\perp}$.  
The distribution 
amplitude $\varphi_{\pi} (x)$  can be treated as the integral of the 
wave function $\Psi(x, k_{\perp})$  over $k_{\perp}$ (cf. \cite{bl80}):
\begin{equation}
\varphi_{\pi}(x) = {{\sqrt{6}}\over{(2\pi)^3}} 
\int \Psi (x,k_{\perp}) \, d^2k_{\perp} \, . 
  \label{66} \end{equation}
For $\Psi^{soft}(x, k_{\perp})$, this integral  
perfectly converges. However, 
 the perturbative $1/k_{\perp}^2$-tail generates
 logarithmic   divergences. Hence, 
one should supplement this definition 
by some regularization  procedure specified 
by a cut-off parameter $\mu$:
$\varphi_{\pi}(x) \to \varphi_{\pi}(x,\mu)$.
The ``cut-off'' should be understood in a broad sense. 
It may be imposed    literally
$k_{\perp}^2 < \mu^2$ or  one can use
 more gentle   procedures based, say,  on  dimensional 
regularization.  In other words, 
$\varphi_{\pi}(x)$ is a scheme-dependent object:
$\varphi_{\pi}(x) \to \varphi_{\pi}^{(S)}(x) $. 
The choice of a specific scheme $S$ is a matter of   convenience.
In particular,   the Fourier transform 
\begin{equation}
\widetilde  \Psi(x,b) = \frac1{(2 \pi)^2} 
\int e^{-i k_{\perp} b_{\perp} } 
 \Psi (x,k_{\perp}) \, 
d^2 k_{\perp}  
\label{67}   
 \end{equation}
to the impact parameter representation
can also be treated (at least,
for small $b$)\footnote{
The basic difference between $\varphi_{\pi}(x;1/b)$
and $\widetilde  \Psi(x,b)$ is that 
$\int_0^1 \varphi_{\pi}(x;\mu)$
is given by the same constant $f_{\pi}$ for any $\mu$ 
while $\int_0^1 \widetilde  \Psi(x,b) dx$ in general depends on $b$.}
 as a regularization scheme 
for  the integral defining the distribution amplitude:
\begin{equation}
\varphi_{\pi}^{(F)}(x;\mu =1/b)  =   {{\sqrt{6}}\over{2\pi}} \, 
\widetilde  \Psi(x,b) \quad ; \quad b  \to 0 \, .
\label{68} \end{equation}
This observation suggests the  extrapolation of 
the   convolution
formula into the nonperturbative region by  substituting
$\varphi (x; 1/b)$ in Eq.(\ref{65}) 
 by the $b$-space wave function $\widetilde \Psi(x,b)$  (see  ref.\cite{kroll}). 
Since the $k_{\perp}$-effects are only essential 
when $x Q^2$ ($i.e., \, x$) is small, one can 
either use  the original combination $\sqrt{x \bar x }Qb$ 
in the argument of the  Born term  $K_0(\sqrt{x \bar x } Qb)$ 
or substitute it by  $\sqrt{x }Qb$. In particular, a modified 
version of the convolution
formula (\ref{65}) written in the 
$k_{\perp}$-representation
\begin{equation}
 F_{\gamma^*\gamma \pi^0}(Q^2) = 
\frac1{\pi^2 \sqrt{6}}
\int_0^1 dx \,  \int \, 
\frac{\Psi (x,k_{\perp})}{ xQ^2 + k_{\perp}^2 } \,  d^2k_{\perp} \, , 
\label{69} \end{equation}
 is the starting
point of the analysis by Jakob $et \ al.$\cite{kroll}. 
In  this expression,  a simpler form  $xQ^2 + k_{\perp}^2$ is used for 
the modified denominator 
of the ``hard'' quark propagator instead of the combination 
$xQ^2 + k_{\perp}^2/\bar x$ which appears in our Eq.(\ref{36}).
However,  since the difference is proportional to 
$k_{\perp}^2$ and vanishes for $x=0$,     the two 
forms  have essentially the same footing.  
 As a model for $\Psi (x,k_{\perp})$, Jakob $et \ al.$\cite{kroll}
use the ansatz \cite{bhl}  with the 
exponential dependence on the combination
$k_{\perp}^2/x\bar x$ (or Gaussian dependence on $k_{\perp}$). We  write 
it  in a   form similar to that used in ref.\cite{kroll}:
\begin{equation}
\Psi^{(G)}  (x,k_{\perp})  = \frac{4 \pi^2}{\sigma \sqrt{6} } 
\, \frac{\varphi_{\pi}(x)}{x \bar x}\,
 \exp \left (- \frac{k_{\perp}^2 }{2\sigma x \bar x} \right )
 \,  , 
\label{70} \end{equation}
where  $\sigma$ is the
width parameter and $\varphi_{\pi}(x)$ is 
the desired  pion distribution amplitude\footnote{In  the original 
model \cite{bhl} $k_{\perp}^2$ appears in the combination
$k_{\perp}^2 + M_q^2$ where $M_q $
is  the constituent quark mass.  
As a result, the distribution amplitude
$\varphi_{\pi}(x)$ is exponentially  suppressed like  
$\exp[-M_q^2/2 \sigma x \bar x]$ 
in the end-point regions.   Jakob $et \ al.$, however,   
follow Chibisov  and Zhitnitsky \cite{chizhit}    
who insist that the constituent quark mass $M_q$ 
should not appear in QCD-motivated models for 
$\Psi  (x,k_{\perp})$.
In particular, $M_q$ does not appear in the  model wave function 
$\Psi^{(LD)}  (x,k_{\perp})$ \cite{apa95} based on local
quark-hadron duality (see Section V below):
only the current quark masses $m_q$ (usually set to zero
for $u$ and $d$ quarks) 
are present in QCD Feynman integrals.}.   
In the $b_{\perp}$-representation, the model wave function is
\begin{equation}
\widetilde \Psi^{(G)}  (x,b_{\perp})
 = \frac{2 \pi}{ \sqrt{6} } \,  \varphi_{\pi}(x) 
\exp \left (- \frac12 \, b_{\perp}^2  \sigma x \bar x \right )
 \,  .
\label{70A} \end{equation}
The  model is restricted by  two conditions taken from ref.\cite{bhl}.
First, the  two-body Fock component 
of the pion light-cone wave function $\Psi (x,k_{\perp})$ 
is required to satisfy the constraint 
\begin{equation}
\int_0^1 dx \,  \int \, \Psi (x,k_{\perp}) \, \frac{ d^2 k_{\perp}}{16\pi^3}  = 
\frac{f_\pi}{2\sqrt{6}}
  \label{71} \end{equation}
imposed by the  $\pi \to \mu \nu$ rate.
This gives the usual normalization condition for the pion DA
\begin{equation}
\int_0^1 \varphi_{\pi}(x) \, dx = f_\pi \, .
\label{72} \end{equation}
The second condition  specifies the value of 
the $x$-integral of  $\Psi (x,k_{\perp})$ 
at zero transverse momentum 
\begin{equation}
\int_0^1 \,  \Psi(x,k_{\perp}=0) \, dx\, = \frac{\sqrt{6}}{f_\pi}.
 \label{73} \end{equation}
For the model ansatz (\ref{70}), this condition results in the 
following constraint for the $I_0$-integral  
\begin{equation} 
 I_0  \equiv \frac1{f_\pi} \int_0^1 
 \varphi_{\pi}(x) \frac{dx}{x} = \frac{ 3 \sigma}{s_0} \, .
 \label{74} \end{equation}
In obtaining Eq.(\ref{74}), we incorporated the symmetry 
property $ \varphi_{\pi}(x) =\varphi_{\pi}(\bar x)$  
of the pion DA and used again the notation 
$s_0$  for the important combination 
$4 \pi^2 f_\pi^2$. 
Since $I_0^{as} =3$ and $I_0^{CZ} =5$, the width parameters are
$\sigma^{as} = s_0 \approx 0.67\,  GeV^2$ and 
$\sigma^{CZ} = \frac{5}{3} s_0 \approx 1.11 \, GeV^2$.

In the form (\ref{73}), the
 second condition  was derived  in ref.\cite{bhl}
from the requirement that the  
$ \pi^0 \to \gamma \gamma$  decay  rate 
(or, what is the same, $F_{\gamma^*\gamma \pi^0}(Q^2=0$) ) 
calculated within the light-cone approach 
coincides with that given  by the axial anomaly. 
It is easy to see, however, that  in the  $Q^2 \to 0$   limit, 
the $k_{\perp}$-integral in Eq.(\ref{69}) logarithmically
diverges in the  small-$k_{\perp}$  region 
for any function which is nonvanishing 
at $k_{\perp}=0$. 
Note, 
that $\Psi (x, k_{\perp}=0)$  cannot vanish
if we wish to satisfy  the condition (\ref{73}).
Rather ironically,   the condition which 
presumably should 
secure the correct value for $F_{\gamma^*\gamma \pi^0}(Q^2)$ 
at $Q^2=0$  guarantees instead  that the extrapolation  
formula diverges  at that point. 
This gives a clear warning   that one should be very careful using 
 the simplest  extrapolation:
 it is  difficult to judge {\it a priori} how reliably 
the formula  failing  for $Q^2=0$  models the subasymptotic effects
for moderate $Q^2$. 
The authors of ref.\cite{kroll} also include the Sudakov 
exponential in which they take a symmetric combination
$s(\bar x Qb) + s( x Qb)$. As noted earlier,
our  one-loop calculation in Sect.III$B$ shows that 
for $F_{\gamma^*\gamma \pi^0} (Q^2)$ one should use 
$ s( \sqrt{x} Qb)$ instead of $s( x Qb)$.
Our final  observation  is that expanding Eq.(\ref{69}) 
in $k_{\perp}^2/Q^2$ one would get an infinite 
 series of power corrections under the $x$-integral. 
According to our general result, 
the handbag diagram should  not produce a chain of higher-twist
contributions.  Hence, the extrapolation formula cannot
be interpreted simply as a transverse-momentum-corrected expression 
for   the handbag diagram.

\subsection{Transverse momentum in the light-cone formalism}

Another  attempt to model the subasymptotic corrections
was made in ref.\cite{huang}. It is based on the 
Brodsky-Lepage    formula \cite{bl80} 
for the two-body ($i.e., \, \bar qq $) contribution
to the $\gamma^*\gamma \pi^0$
form factor in the light-cone formalism:
\begin{equation}
(\epsilon_{\perp} \times q_{\perp}) F^{\bar qq}_{\gamma^*\gamma \pi^0} 
(Q^2) = {{1}\over {\pi^2 \sqrt{6}}} 
\int_0^1 dx \int  
\frac{(\epsilon_{\perp} \times (xq_{\perp}+k_{\perp})) }
{ (xq_{\perp}+k_{\perp})^2- i \epsilon} \, \Psi(x,k_{\perp}) \,  d^2 k_{\perp} \, .
  \label{75} \end{equation}
Here, $q_{\perp}$ is a two-dimensional vector in the transverse plane
satisfying $q_{\perp}^2=Q^2$,   $\epsilon_{\perp}$ is a vector orthogonal to 
 $q_{\perp}$ and also lying in 
the transverse plane \cite{bl80} and the cross denotes the vector product.
Again, the wave function is chosen  in the Gaussian form (\ref{70}) 
satisfying  the  constraints (\ref{71}) and (\ref{73})
\footnote{As  emphasized recently by Kroll \cite{kroll2},
Cao {\it et al.}  use constituent quark masses $M_q \sim 330 \, MeV$
which produces  a strong exponential suppression $\exp [-M_q^2/2 \sigma x \bar x]$
of  the end-point regions. As a result, the $I$-integral 
for the DA corresponding to their ``CZ'' model
is $3.71$ rather than 5, $i.e.,$  despite  zero at $x =1/2$,
 such a model gives a rather narrow  DA, which is closer in this
sense to the asymptotic DA rather than to the original CZ  one. }.
 Though  the integrand of  Eq.(\ref{75}) looks   rather singular, there 
are  no problems 
with the convergence of 
the $k_{\perp}$-integral     in the 
$q_{\perp} \to 0$ limit. The  result is finite, since 
\begin{equation}
\left. \frac{ q_{\perp}^{\alpha} +k_{\perp}^{\alpha}}
{ (q_{\perp}+k_{\perp})^2- i \epsilon} \, \right |_{q_{\perp} \to 0}  
= \pi \, \delta^2 (k_{\perp}) \, q_{\perp}^{\alpha} 
\label{76} \end{equation}
for  any test function $\Psi(x,k_{\perp})$  
which depends on $k_{\perp}$ through $k_{\perp}^2$.
Because of the $\delta^2(k_{\perp})$-function, 
the $Q^2=0$ result is determined   by the  wave function
at zero transverse momentum.

In  ref.\cite{huang}, it is 
claimed  that the $k_{\perp}/Q$ expansion of 
Eq.(\ref{75}) produces  large ``higher-twist''  
corrections to the leading-twist  
result. 
In fact, when  $\Psi(x,k_{\perp})$ has an 
 exponential   $k_{\perp}^2$-dependence,
it is trivial to calculate
the $k_{\perp}$-integral  explicitly 
\begin{equation}
F^{\bar qq}_{\gamma^*\gamma \pi^0}(Q^2) = 
\frac{4 \pi}{3} \int_0^1 \frac{\varphi_{\pi}(x)}{x Q^2} \left [ 1- \exp \left 
( -\frac{xQ^2}{ 2\bar x \sigma } \right ) \right ] dx 
\label{77} \end{equation} 
to  see that 
the correction term in the integrand of Eq.(\ref{77}) 
has an exponentially decreasing
rather than a power behavior for large $Q^2$.  
This result  agrees with 
our general statement that
the handbag diagram 
 contains no higher-twist contributions.
Our  analysis works in this case since 
the Brodsky-Lepage formula (\ref{75})
 corresponds to  the  handbag contribution
written in the light-cone  variables  without any 
approximation.  
Just like in the covariant treatment, 
the naively expected series of power  
corrections $(\langle k_{\perp}^2 \rangle/Q^2)^n$
does not appear 
because the expansion of 
\begin{equation} 
 {{xq_{\perp}+k_{\perp}}\over{(xq_{\perp}+k_{\perp})^2}}
\label{78} \end{equation}
contains only traceless combinations.
Indeed,  multiplying
 (\ref{78})  by $q_{\perp}/Q^2$
and defining 
$(k_{\perp} q_{\perp})= |k_{\perp}| Q \cos \phi$,
we obtain 
\begin{eqnarray} 
&& \lefteqn{\left ( \frac1{Q^2} \right ) \frac{xQ^2+
|k_{\perp}|Q\cos \phi}{x^2 Q^2+ 2 x |k_{\perp}|Q\cos \phi + k_{\perp}^2}  = 
\label{79} } \\ && \frac{1}{xQ^2}\left \{ {\theta(|k_{\perp}|<xQ)} + \sum_{n=1}^{\infty}
(-1)^n \left [  \left (\frac{|k_{\perp}|}{xQ} \right )^{n} 
{\theta(|k_{\perp}|<xQ)} -
 \left (\frac{xQ}{|k_{\perp}|} \right )^{n} {\theta(|k_{\perp}|>xQ)} \right ]
 \cos (n \phi) \right \}.
\nonumber 
 \end{eqnarray}
For a wave function  $\Psi(x,k_{\perp})$  depending     on $k_{\perp}$
 through $k_{\perp}^2$ only,  
all the oscillating  terms proportional to 
$\cos (n \phi)$  ($i.e.,$ to Chebyshev polynomials
$T_n(\cos \phi)$ corresponding to traceless
combinations in two dimensions)
vanish after the angular integration. 
Only  the  $n=0$ term written outside the sum over $n$
gives a nonzero result. 
Hence,  for   the wave functions of $\Psi(x,k_{\perp})= \psi(x,k_{\perp}^2)$
type,  we can write 
\begin{equation}
F^{\bar qq}_{\gamma^*\gamma \pi^0} 
(Q^2) = {{2}\over {\pi \sqrt{6}}} 
\int_0^1 \frac{dx}{xQ^2} \int_0^{xQ}  
\, \psi(x,k_{\perp}^2) \, 
k_{\perp} d k_{\perp}
\, .
  \label{75A} \end{equation}
This  means that the leading $1/xQ^2$ term in Eq. (\ref{77}) comes from 
the  integral over all $k_{\perp}$'s  while the exponential correction 
 appears because the integration region in (\ref{75A})
is restricted by $k_{\perp}<xQ$. 
Another subtlety is that the $Q^2=0$ value 
$$F^{\bar qq}(Q^2=0)=\frac1{2 \pi f_\pi}$$
dictated  by eqs.(\ref{73}) and (\ref{76})
(and manifest in Eq.(\ref{77}) )
 gives only a  half of what is needed to get the correct
$\pi^0 \to \gamma \gamma$ rate (\ref{58}) . 
As explained in ref.\cite{bhl}, the other half comes from 
the term which can be interpreted as the
contribution of the $\bar qq \gamma$ Fock component of the 
pion wave function.  In a formal  pQCD diagrammatics,  
this contribution  is represented  by graphs containing the 
gluons coupling  to the 
quark line between  the photon vertices.
For high $Q^2$, such diagrams correspond to higher-twist 
corrections associated with  the $\bar q G \ldots G q$ operators.
In this sense, the result of ref.\cite{bhl} 
is equivalent  to a
nonperturbative constraint 
on the $Q^2 \to 0$ limit of such contributions. 
One can  expect that 
the $\bar qq \gamma$ contribution  decreases as $1/Q^4$ or faster
for large $Q^2$  since it contains higher twists only.  
Interpretation of this contribution 
in terms of the  $\bar qq \gamma$ Fock component 
is restricted to the case of  real $\gamma$:  
  ref.\cite{bhl} gives no expression 
beyond the $Q^2=0$ point. In ref.\cite{huang} 
this contribution is not included.
However,  if   the terms which double
the result for $Q^2=0$ are not included, 
it is premature  to make specific  
 quantitative statements about  the size of 
subasymptotic corrections in the region
of moderate $Q^2$.  

We may also wonder {\it why}  the formulas (\ref{69}) and (\ref{75}) 
corresponding to two  attempts to include the
primordial transverse momentum have such a strikingly different
analytic structure. In particular, the denominator of the 
integrand of Eq.(\ref{75}) 
vanishes  for $k_{\perp} = -xq_{\perp}$ while that of 
Eq.(\ref{69})  is finite  for all $k_{\perp}$ 
provided that $q_{\perp} \neq 0$.
The   answer is very simple: the two expressions imply
two  different definitions of 
what is longitudinal and what is transverse.
Eq. (\ref{69}) is based on the Sudakov decomposition in which
the momentum $q_1$ of the real photon has only the  light-cone ``plus''
component while the momentum
$p$ of the  pion has only  the light-cone ``minus''
component. As a result,  the   momentum transfer $q_2=p-q_1$
in the Sudakov variables  is purely longitudinal
and has both plus and minus components, with $q_2^2 = -2(q_1 p)$.  
On the other hand, the Brodsky-Lepage formula corresponds to 
 the infinite
momentum frame in which the plus components of $q_1$ and $p$ coincide.
The plus component of the momentum transfer $q_2$ vanishes
in this frame,
but $q_2$  has a nonzero transverse component $q_{\perp}$, with $|q_{\perp}|=Q$
or $q_2^2 = -q_{\perp}^2$. 
Evidently, the two frames cannot be obtained from one another
by a boost. Furthermore, one should  not 
expect a diagram by diagram correspondence between the
two approaches. The main purpose of imposing   the  
requirement $q_2^+=0$ in the 
light-cone approach is to  avoid the  $Z$-graphs.
 However,   in Sudakov variables (and in any approach in which 
$q_2$ has a non-zero plus component) 
  the  $Z$-graphs should be added to reproduce 
the light-cone result (cf.\cite{sawicki}).

Both the  approaches \cite{kroll,huang} discussed above
fail to reproduce  the $Q^2=0$ value  corresponding to
 the axial anomaly. Our point of view is that
complying  with the  anomaly constraint 
should be a minimal requirement for any model 
of subasymptotic effects in the $\gamma^* \gamma \pi^0$ form factor.
 A maximalist attitude  is 
   that such a fundamental   constraint 
should be satisfied automatically rather than imposed 
as an external condition.
This can be only realized in  an approach which is directly related
to QCD and produces anomaly as a consequence of  QCD dynamics.

\section{Quark-hadron duality and effective wave function}

\subsection{QCD sum rule calculation of $f_{\pi}$ and local duality}

QCD sum rules provide us with the 
 approach which deals   both with perturbative and nonperturbative 
aspects of QCD. 
The basic idea of the QCD sum rule approach \cite{svz}
is the quark-hadron  duality,
$i.e.,$ the possibility to describe one and the same 
object  in terms of either quark/gluon  or hadronic fields.
To get information about the pion, 
the QCD sum rule practitioners  usually 
  analyze  correlators involving the axial current.
In particular, to calculate 
$f_{\pi}$ one should  consider  the $p_{\mu}p_{\nu}$-part  of the 
correlator of two axial currents:
\begin{equation} 
\Pi^{\mu\nu}(p) =
i \int e^{ipx}  \langle 0 | T \,(j_{5\mu}(x)\,j_{5\nu}(0)
\,)|\,0\rangle\, d^4 x = 
p_{\mu}p_{\nu}\Pi_2(p^2)-g_{\mu\nu}\Pi_1(p^2).
\label{80} \end{equation}
The dispersion relation
\begin{equation}
 \Pi_2(p^2)= \frac1{\pi}\int_0^{\infty}\frac{ \rho(s)}{s-p^2}ds + 
``subtractions" 
\label{81} \end{equation}
represents $\Pi_2(p^2)$ as an integral  over hadronic spectrum 
with the spectral
density $\rho^{hadron}(s)$  determined by projections 
\begin{equation}
\langle 0 | j_{5\mu} (0) |\pi;  P \rangle = i f_{\pi} P_{\mu},
\label{82} \end{equation}
$etc.,$ of the axial current onto
hadronic states
\begin{equation}
\rho^{hadron}(s) = \pi f_{\pi}^2 \delta(s-m_{\pi}^2) + \pi f_{A_1}^2
\delta(s-m_{A_1}^2)  + ``higher \ \  states" 
\label{83} \end{equation}
( $f_{\pi}^{\exp} \approx 130.7 \, MeV$ in our normalization).
On the other hand, when the probing virtuality  is negative and large,
one can use the operator product expansion 
\begin{equation}
\Pi_2(p^2) = \Pi_2^{quark}(p^2) + \frac{A}{p^4} \langle \alpha_s GG \rangle 
+ \frac{B}{p^6} \alpha_s \langle \bar qq \rangle^2  + \ldots
\label{84} \end{equation}
where $ \Pi_2^{quark} (p^2)$ is the 
perturbative  version  of   $\Pi_2 (p^2)$ given by a sum of 
pQCD Feynman diagrams while the condensate terms 
$\langle GG \rangle$,  $\langle \bar qq \rangle$, $etc.$ 
(with perturbatively calculable coefficients
$A, B,$  see Eq.(\ref{89}) below), 
describe/parametrize  the  nontrivial structure of the QCD vacuum.
For the quark amplitude $\Pi_2^{quark}(p^2)$,  one can also write down 
the  dispersion relation  (\ref{81}),  with $\rho(s)$ 
substituted by its perturbative analog $\rho^{quark}(s)$:
\begin{equation}
\rho^{quark}(s)= \frac1{4 \pi} \left ( 1 + \frac{\alpha_s}{\pi} 
+ \ldots \right ) 
\label{85} \end{equation}
(we neglect light quark masses). Hence, for large $-p^2$, one can write 
\begin{equation}
\frac1{\pi} \int_0^{\infty} {{\rho^{hadron}(s) - \rho^{quark}(s)}\over
{s-p^2}} ds \ = \frac{A}{p^4} \langle \alpha_s GG \rangle 
+ \frac{B}{p^6} \alpha_s\langle \bar qq \rangle^2  + \ldots \ .
\label{86} \end{equation}
This expression essentially states that the condensate 
terms describe  the difference between  the 
quark and hadron spectra. 
At this point, using the known values of the condensates,
one can try to construct  a model for the hadronic spectrum.
In the axial-current channel, one has 
an infinitely   narrow pion peak 
$\rho_{\pi} = \pi f_{\pi}^2 \delta(s-m_{\pi}^2)$, 
a rather wide  peak at 
$s \approx 1.6 \, GeV^2$ corresponding to  $A_1$ and then   a 
 ``continuum'' at higher energies. The simplest approximation 
is to treat  $A_1$ also as a part  of the continuum, 
$i.e.,$ to use  the model 
\begin{equation}
\rho^{hadron}(s) \approx \pi f_{\pi}^2 \delta(s-m_{\pi}^2) + \rho^{quark}(s) \, 
 \theta(s \geq s_0), 
\label{87} \end{equation}
in which   all the higher resonances including the  $A_1$ 
are approximated  by the quark  spectral density  starting at some 
effective threshold $s_0$.  
Neglecting the pion mass and  requiring the best agreement between the two sides 
of the resulting  sum rule
\begin{equation}
{{f_{\pi}^2}\over{p^2}} = \frac1{\pi} \int_0^{s_0} {{\rho^{quark}(s)}\over
{s-p^2}} ds \ + \frac{A}{p^4} \alpha_s \langle GG \rangle 
+ \frac{B}{p^6} \alpha_s \langle \bar qq \rangle^2 + \ldots \ 
\label{88} \end{equation}
in the region of large $p^2$, 
we can fit the remaining parameters  $f_{\pi}$ and  $s_0$ 
characterizing the model spectrum.
In practice, the more convenient SVZ-borelized version \cite{svz}
of this sum rule 
\begin{equation} 
{f_{\pi}^2} = \frac1{\pi} \int_0^{s_0} \rho^{quark}(s) e^{-s/M^2} {ds}
 \  +\frac{\alpha_s\langle GG\rangle}{12\pi M^2}
		  +\frac{176}{81}\frac{\pi\alpha_s\langle\bar qq\rangle^2}{M^4}
		    + \ldots
  \label{89} \end{equation}
is used for actual fitting.
Using the standard values for the condensates
$\langle GG \rangle$, $\langle \bar qq \rangle^2$, the scale 
 $s_0$ is  adjusted to get an (almost) constant  result
for the rhs of Eq.(\ref{89}) 
starting with the minimal possible value of the SVZ-Borel parameter $M^2$.  
The magnitude  of   $f_{\pi}$ extracted in this way,  is very close
to its  experimental value $f_{\pi}^{exp} \approx 130 \, MeV.$

Of course, changing the  values of the condensates, 
one would get the best stability for a different magnitude 
of  the effective threshold 
$s_0$, and the resulting value of  $f_{\pi}$ would also change.
There exists  an evident  correlation  
between the values of $f_{\pi}$ and $s_0$ since, in the 
$M^2 \to \infty$ limit, the sum rule reduces to the local duality relation 
\begin{equation}
f_{\pi}^2 = \frac1{\pi} \int_0^{s_0} \rho^{quark}(s) \, ds.
  \label{90} \end{equation}
Using the explicit lowest-order expression 
$\rho^{quark}_0(s) = 1/4\pi$, we get 
\begin{equation}
s_0 = 4\pi^2 f_{\pi}^2.
 \label{91} \end{equation}
Note that $s_0 = 4\pi^2 f_{\pi}^2$ coincides 
with  the combination which appears 
in the Brodsky-Lepage interpolation formula 
(\ref{60}).

\subsection{Quark-hadron duality for the 
$F_{\gamma^* \gamma^* \pi^0} \left(Q^2\right)$
form factor}

    Information about   the 
$\gamma^* \gamma^* \to \pi^0$ form factor   
can be extracted from   the three-point correlation function   \cite{NeRa83}
\begin{equation}
{\cal{F}}_{\alpha\mu\nu}(q_1,q_2)= \frac{4\pi}{i \sqrt{2}} 
\int d^4x\,d^4y\ e^{-iq_1 x- iq_2 y}
\langle 0 |T\left\{J_{\mu}(x)\,J_{\nu}(y)\,j_{5 \alpha}(0)\right\}| 0 \rangle
  \label{92} \end{equation}
calculated in the region where all the  
virtualities \ $q_1^2 \equiv - q^2 ,q_2^2\equiv -Q^2$ and 
$p^2\!=\!(q_1+q_2)^2$ are spacelike.

The form factor $F_{\gamma^* \gamma^* \pi^0}(q^2,Q^2)$ 
appears in 
the invariant  amplitude $F \left(p^2,q^2,Q^2\right)$ corresponding to the 
tensor structure ${\epsilon}_{{\mu}{\nu}{\rho}{\sigma}} 
p_{\alpha} q_1^{\rho}q_2^{\sigma}$.
The dispersion relation 
for the three-point  amplitude 
\begin{equation}
F\left(p^2, q^2,Q^2 \right)={1\over{\pi}}\int_0^{\infty}
\frac{{\rho}\left(s,q^2,Q^2\right)}{s-p^2}\,ds
+ \ ``subtractions" 
  \label{93} \end{equation}
specifies the relevant spectral density ${\rho}\left(s,q^2,Q^2\right)$.
For the  hadronic spectrum we assume again the ``first resonance plus
perturbative continuum'' ansatz
\begin{equation}
{\rho}^{hadron}\left(s,q^2,Q^2\right) = \pi f_{\pi} 
F_{\gamma^* \gamma^* \pi^0}(q^2,Q^2)
\delta(s-m_{\pi}^2) +  \theta(s>s_0) \, \rho^{quark}(s,q^2,Q^2)\, .
  \label{94} \end{equation}
The lowest-order perturbative spectral density ${\rho}^{quark}(s,q^2,Q^2)$
is  given by  the  Feynman parameter representation
\begin{equation} 
\rho^{quark}(s,q^2,Q^2)=2 \int_0^1 
\delta \left ( s - \frac{q^2x_1x_3+Q^2x_2x_3}{x_1x_2} \right ) \
\delta \left (1-\sum_{i=1}^3 x_i \right ) \
dx_1dx_2dx_3 \, . 
 \label{95} \end{equation}
 Scaling  the integration variables: 
$x_1+x_2 =  y$, $x_2 =  xy$, $x_1 =(1-x)y \equiv \bar x y$ 
and taking trivial integrals over $x_3$ and $y$, we get 
\begin{equation}
\rho^{quark}(s,q^2,Q^2)=2\int_0^1 \frac{x\bar{x}(xQ^2+ \bar x q^2)^2}
{[s{x}\bar{x}+xQ^2+ \bar x q^2]^3} \,dx  \, .
  \label{96} \end{equation}
The variable $x$  
here can be treated as  the light-cone fraction of the 
pion momentum $p$ carried by one of the quarks.
In particular, the denominator of the
integrand in Eq.(\ref{96})  is related to that  of the  
hard quark propagator: $(q_1-xp)^2 = -(xQ^2+ \bar x q^2+ s{x}\bar{x})$.

Putting one photon on shell,  $q^2=0$, 
we can easily calculate the $x$-integral:
\begin{equation}
\rho^{quark}(s,q^2=0,\, Q^2) = 2\int_0^1 
\frac{x\bar{x}(xQ^2)^2}
{[s{x}\bar{x}+xQ^2]^3} \,dx  \, = 
{{Q^2}\over{(s+Q^2)^2}} \, . 
   \label{96A} \end{equation}
This result explicitly shows that if 
the larger virtuality $Q^2$ also  tends to 
 zero, the spectral density $\rho^{quark}(s,Q^2)$ 
becomes narrower and higher, approaching  $\delta(s)$ in the
$Q^2 \to 0$ limit (cf. \cite{DZ}).
Thus,  the perturbative triangle diagram
dictates that two real photons can produce only 
a single  massless pseudoscalar state: 
there are no other states in  the spectrum  of final hadrons
(cf. \cite{consist}).
As $Q^2$ increases, the spectral function broadens,
$i.e.,$ higher states can  also be  produced.

A detailed study of the QCD sum rule for 
the $F_{\gamma^* \gamma \pi^0}(Q^2)$ form factor
was  performed in refs.\cite{pl,rr}.  
The results of this  investigation 
are rather close to those based on  
the simple local quark-hadron duality ansatz:
\begin{equation}
 F_{\gamma^* \gamma \pi^0}^{LD}(Q^2)
= \frac1{\pi f_{\pi}} \int_0^{s_0} \rho^{quark}(s, Q^2)\, ds \, .
\label{97} \end{equation}
Using the explicit expression for $\rho^{quark}(s, Q^2)$, we can write
\begin{equation}
F_{\gamma^* \gamma\pi^0}^{LD}(Q^2)
=  \frac2{\pi f_{\pi}} 
\int_0^1 dx \int_0^{s_0}  \frac{x\bar{x}(xQ^2)^2}
{[s{x}\bar{x}+xQ^2
]^3}  \, ds  = \frac1{\pi f_{\pi} (1+Q^2/s_0)} \, . 
\label{98} \end{equation}
This result coincides with the Brodsky-Lepage interpolation formula
(\ref{60}).

\subsection{Effective wave function}

The formulas  based on  the
local quark-hadron duality prescription  
can be interpreted in terms 
of the effective two-body light-cone wave function \cite{apa95}.
Consider the lowest-order perturbative spectral density 
for the two-point correlator.  It can be written as 
the Cutkosky-cut quark loop integral 
\begin{equation}
\rho^{quark}(s) = \frac{3}{2\pi^2} \int 
\frac{k_+}{p_+} \left ( 1- \frac{k_+}{p_+}  \right )
\, \theta(k_+) \,  \delta \left (k^2 \right ) 
\, \theta(p_+ - k_+) \, \delta
\left ( (p-k)^2 \right ) d^4k
  \label{100} \end{equation}
where $s\equiv p^2$. 
Introducing the light-cone variables for $p$ and $k$:
$$p= \{p_+ \equiv P, p_- = s/P, p_{\perp} =0 \} \  ;  \ 
k = \{k_+\equiv xP, k_-, k_{\perp} \}$$
and integrating over $k_-$, we get
\begin{equation}
\rho^{quark}(s) = \frac{3}{2\pi^2} \int_0^1 dx \int 
 \delta \left (s- \frac{k_{\perp}^2}{x \bar x} \right )
 \, d^2 k_{\perp} .
\label{101}    \end{equation}
The delta-function  here   expresses the 
fact that 
the light-cone combination 
${{k_{\perp}^2}/{x \bar x}}$  coincides with $s\equiv p^2$, 
the invariant mass of the $\bar q q$ pair. 
Substituting this expression  for
 $\rho^{quark}(s)$ into the local duality formula (\ref{90}), 
we obtain the following representation for $f_{\pi}^2$
\begin{equation}
f_{\pi}^2 =  \frac{3}{2\pi^3} \int_0^1 dx \int 
\theta \left (k_{\perp}^2\leq  x\bar x s_0 \right ) \, d^2 k_{\perp} \, .
 \label{102} \end{equation}
It has the structure similar to the expression for $f_{\pi}$ 
in the light-cone formalism \cite{bl80} (cf. (\ref{71}))
\begin{equation}
f_{\pi} = \sqrt{6} \, 
\int_0^1 dx \int \Psi(x,k_{\perp}) \, \frac{d^2 k_{\perp}}
{8 \pi^3} \, 
 .
 \label{103} \end{equation}
To cast  the local duality result
(\ref{102}) into the form of Eq.(\ref{103}),
we introduce the ``local duality''
wave function for the pion:
 \begin{equation} 
\Psi^{LD}(x,k_{\perp}) = \frac{2\sqrt{6}}{f_\pi} \,
\theta(k_{\perp}^2\leq  x\bar x s_0) \ .
  \label{104} \end{equation}
 The specific form dictated by the local duality
implies that $\Psi^{LD}(x,k_{\perp})$ 
simply imposes a sharp cut-off at $ k_{\perp}^2x\bar x = s_0$.
In the  $b_{\perp}$-space, the   effective 
wave function can be written as 
\begin{equation}
\widetilde \Psi^{LD} (x, b_{\perp}) = {{\sqrt{6}}\over{\pi f_{\pi} b_{\perp}}}  
\sqrt{x \bar x s_0} \, 
J_1(b_{\perp} \sqrt{x \bar x  s_0 }),
\label{eq:bWF}
\end{equation}
where $J_1(z)$ is the Bessel function.

\subsection{Effective wave function and 
$F_{\gamma^*\gamma \pi^0} (Q^2)$ form factor} 

Consider now the local duality expression (\ref{98}) for 
$F_{\gamma^* \gamma \pi^0} ( Q^2)$. 
Replacing   $s$, 
the invariant mass of the $\bar qq$ pair, 
 by its  light-cone equivalent 
$k_{\perp}^2/ x\bar x $, we get  
$F_{\gamma^* \gamma \pi^0}^{LD}(Q^2)$
as an integral  over 
the longitudinal momentum fraction 
$x$ and the  transverse momentum $k_{\perp}$:
\begin{eqnarray}
F_{\gamma^* \gamma \pi^0}^{LD}(Q^2) = 
\frac2{\pi^2 f_{\pi}} 
\int_0^1 dx \, \int  \, 
\frac{(xQ^2)^2}{ (xQ^2+ k_{\perp}^2)^3} \,
\theta(k_{\perp}^2\leq  x\bar x s_0)\, d^2k_{\perp} \,.
 \label{105} \end{eqnarray}
Now, introducing the effective wave function $\Psi^{LD}(x,k_{\perp})$
given by (\ref{104}), 
we  write $F^{LD}\left( Q^2\right)$  in the ``light-cone form'':
\begin{equation}
 F_{\gamma^* \gamma \pi^0}^{LD}(Q^2) = \frac1{\pi^2 \sqrt{6}}
\int_0^1 dx \, \int  \,
\frac{(xQ^2)^2}{ (xQ^2 +k_{\perp}^2)^3} \,
\Psi^{LD}(x,k_{\perp}) \,  d^2k_{\perp} . 
  \label{106} \end{equation}
In  the impact parameter representation, this formula
looks like 
\begin{equation}
F_{\gamma^*\gamma \pi^0}^{LD}(Q^2)  =
 \frac1{2 \pi  \sqrt{6}}\,  \int \limits_0^1 dx \int  xQ^2 b^2 
\ K_2 \left (\sqrt{x} b Q   \right )
        \widetilde      \Psi^{LD} (x, b_{\perp})         \, 
d^2 b_{\perp}  . 
\label{35A} \end{equation}
The function  $K_2 \left (\sqrt{x} b Q   \right )$,
where  $K_2(z)$ is the modified Bessel function, 
originates from    the  new version of the Born term written
in the $b$-space 
\begin{equation}
 \tilde B(x;bQ) \equiv 
  \frac{1}{2 \pi}\int   e^{- i k_{\perp} b_{\perp} }
{(x Q^2)^2 \over (x Q^2 + k_{\perp}^2)^3} \,  d^2 k_{\perp}
   =  \frac14 \, xQ^2 b^2   K_2 \left ( \sqrt{x} b Q  \right ) \, .
\label{36A} \end{equation}
Note that $ \tilde B(x;bQ)$ is finite for $b=0$: 
 $ \tilde B(x;0)=1$ while the ``old'' Born term 
 $ B(x; bQ) = \bar x K_0 \left ( \sqrt{x\bar x } b Q  \right ) $ 
 (\ref{36}) has a logarithmic singularity 
at the origin of the $b$-space.
The  expression  (\ref{106}) 
looks similar to the extrapolation formula
(\ref{69}). Furthermore, since
\begin{equation}
\frac{(xQ^2)^2}{ (xQ^2 +k_{\perp}^2)^3} 
= \frac{1}{ xQ^2 +k_{\perp}^2}
- \frac{2 k_{\perp}^2}{ (xQ^2 +k_{\perp}^2)^2} + 
\frac{k_{\perp}^4 }{ (xQ^2 +k_{\perp}^2)^3} \, , 
 \label{106A} \end{equation}
the two $k_{\perp}$-modifications of the hard quark propagator
$1/xQ^2$ differ only by $O(k_{\perp}^2)$ terms 
invisible in the analysis of effects induced by
the $1/k_{\perp}^2$ singularity at small $k_{\perp}$.
However, this difference is very essential when 
one extrapolates into the region of small $Q^2$.
To demonstrate this, let us   analyze Eq.(\ref{106})  in some 
particular limits. 
For real photons, using 
the fact that  
\begin{equation}
\frac{\mu^4}{ (\mu^2+k_{\perp}^2)^3} \to \frac1{2} \,  \delta (k_{\perp}^2)
  \label{107} \end{equation}
in the $\mu^2 \to 0$ limit, 
  we obtain that 
 the $\pi^0 \to \gamma \gamma$
decay rate is determined by the magnitude of the 
$LD$ wave function
at zero transverse momentum:
\begin{equation}
F_{\gamma^* \gamma \pi^0}^{LD}(0) = 
\frac1{2 \pi \sqrt{6}} \int_0^1 \,
\Psi^{LD}(x,k_{\perp}=0)  \, dx .
  \label{108} \end{equation} 
This requirement is similar to that in the
 Brodsky-Lepage formalism. 
However, according to  the explicit form (\ref{104}) 
of $\Psi^{LD}(x,k_{\perp}=0)$,
the integral (\ref{108}) is twice larger than 
the  constraint (\ref{73}) imposed on the valence $\bar qq$
light-cone wave function. 
As a result, the  local duality  formula 
exactly reproduces the $F_{\gamma^* \gamma \pi^0}(0)$ 
value (\ref{58}) dictated by the axial anomaly.
This outcome can be interpreted by saying that
  $\Psi^{LD}(x,k_{\perp})$ is an  
{\it effective}  wave function (cf.\cite{schlumpf})
describing the soft content 
of all $\bar q G \ldots Gq$ 
Fock components of the usual light-cone approach 
(see also \cite{chizhit}). 
Note, that higher-order radiative corrections to  the perturbative
spectral density $\rho^{quark}(s,Q^2)$ 
are explicitly accompanied by the $\alpha_s(\mu_R^2)/\pi$ factors 
per each extra loop. After integration
over the duality interval $0 \leq s \leq s_0$, there are two 
physical  scales: $s_0$ and $Q^2$. 
At low $Q^2$, the duality interval $s_0$ sets the  scale
at the low-momentum end of the $UV$-divergent integrals,
hence,  a natural choice for  
 the normalization scale $\mu_R$ is  $\mu_R^2 \sim s_0$. 
At high $Q^2$, the short-distance dominated parts
of the higher-order corrections  should reproduce 
the pQCD results which suggest  $\mu_R^2 \sim Q^2$
for these terms.  In any case,  suppression by  at least 
$\alpha_s(s_0)/\pi \sim 0.1$ per each extra loop is guaranteed.
Since $s_0 \gg \Lambda^2$,   
 the gluonic corrections to $\rho^{quark}(s,Q^2)$
are suppressed by powers of $\alpha_s(s_0)/\pi \sim 0.1$.
In other words, the  higher-order diagrams 
contributing to $\rho^{quark}(s,Q^2)$ 
correspond to exchange of hard gluons whose  wave lengths 
are larger than $1/\sqrt{s_0}$.

When   $Q^2$ is so large
that the $k_{\perp}^2$-term can be neglected, we get the expression 
\begin{equation}
F_{\gamma^* \gamma \pi^0}^{LD}(Q^2) = \frac1{\pi^2 \sqrt{6}}
\int_0^1 \frac{dx}{xQ^2} \, 
\int  \,\Psi^{LD}(x,k_{\perp}) \, d^2k_{\perp} \, + O(1/Q^4) \,  .
  \label{109} \end{equation}
Identifying the wave function
integrated over the transverse momentum
with  the pion distribution amplitude 
\begin{equation}
\varphi_{\pi}^{LD}(x) \equiv {{\sqrt{6}}\over{(2\pi)^3}} 
\int \Psi^{LD}(x,k_{\perp}) \, d^2k_{\perp} \, = 6 f_{\pi} x(1-x)  , 
\label{110}   \end{equation}
we obtain  the lowest-order pQCD formula (\ref{8})
\begin{equation}
F_{\gamma^* \gamma  \pi^0 }(Q^2)|_{Q^2 \to \infty}  = \frac{4\pi}{3}
\int_0^1 {{\varphi_{\pi}(x)}\over{xQ^2}} \, dx  +O(1/Q^4)
  \label{111} \end{equation}
for the large-$Q^2$  behavior of the $\gamma^* \gamma \to \pi^0$
transition form factor. 

To summarize,    the local duality  formula (\ref{98})
exactly reproduces the  Brodsky-Lepage interpolation (\ref{60})
between the $Q^2=0$ value $1/\pi f_{\pi}$  fixed by the axial anomaly  
and the leading large-$Q^2$ term ${4\pi  f_{\pi}}/{Q^2}$
calculated for the  asymptotic form of the pion distribution amplitude.

The application of the local duality ansatz 
in a general situation when both photons are virtual
was discussed in ref.\cite{apa95}.
The basic formula written in terms of the effective wave function 
is given by 
\begin{equation}
 F_{\gamma^* \gamma^* \pi^0}^{LD}(q^2,Q^2)
= \frac1{\pi f_{\pi}} \int_0^{s_0} \rho^{quark}(s,q^2,\, Q^2)\, ds 
=  \frac2{\pi f_{\pi}} 
\int_0^1 dx \int  \frac{x\bar{x}(xQ^2+ \bar x q^2)^2}
{[k_{\perp}^2+xQ^2
+ \bar x q^2]^3}  \, \Psi^{LD}(x,k_{\perp}) \, d^2k_{\perp} \, .
\label{97A} \end{equation}
For $q^2=Q^2=0$ it satisfies
the anomaly constraint  (\ref{58}), while when  both
$q^2$ and $Q^2$ are large it reduces to  the 
 pQCD formula
\begin{equation}
F_{\gamma^* \gamma^*  \pi^0 }(Q^2)|_{q^2,Q^2 \to \infty}  = \frac{4\pi}{3}
\int_0^1 {{\varphi_{\pi}(x)}\over{xQ^2+\bar x q^2}} \, dx  +O(1/Q^4) \, .
  \label{111A} \end{equation}

\subsection{Extended local duality}

Note,  that the pion distribution amplitude (\ref{110}) 
produced  by  the local duality prescription
coincides with the asymptotic DA. 
To  model wave functions corresponding to  
  DA's different from
$\varphi^{as}_{\pi}(x)$,  we propose to use the sharp 
cut-off analog of the Gaussian model  (\ref{70}):
\begin{equation} 
\Psi^{(LD)}(x, k_{\perp})  = \frac{8 \pi^2}{\sigma \sqrt{6} } 
 \, \frac{\varphi_{\pi}(x)}{x \bar x}\,
\theta \left ( {k_{\perp}^2  \leq  x \bar x \sigma} \right ) \, , 
\label{112} \end{equation}
where $\sigma $ is again the width parameter and $\varphi_{\pi}(x)$ 
the desired DA,  which satisfies the standard 
$f_{\pi}$-normalization constraint
(\ref{72}).  
To guarantee the anomaly result for 
the $\pi^0 \to \gamma \gamma$ rate, 
  we  impose the  following    constraint on the
 $x$-integral of  $\Psi^{(LD)} (x,k_{\perp})$ 
at zero transverse momentum 
\begin{equation}
\int_0^1 \,  \Psi^{(LD)}(x,k_{\perp}=0) \, dx \, = \frac{2\sqrt{6}}{f_\pi} \ .
 \label{113} \end{equation}
Substituting the model ansatz (\ref{112}), we derive from this  
constraint the condition for the standard integral $I_0$ 
\begin{equation} 
 I_0  \equiv \frac1{f_\pi} \int_0^1 
 \varphi_{\pi}(x) \frac{dx}{x} = \frac{ 3 \sigma}{s_0} .
\label{114} \end{equation}
where $s_0$ is  the   basic combination 
$s_0 = 4 \pi^2 f_\pi^2 $. 
Taking  $I_0^{as} =3$ and $I_0^{CZ} =5$, we fix  the width parameters
$\sigma^{as} = s_0 $ and 
$\sigma^{CZ} = \frac{5}{3} s_0 \approx 1.11 \, GeV^2$.
Note,  that in the CZ calculation \cite{cz82}, the duality interval
 was $0.75 \, GeV^2$
 for the zeroth moment of the DA and $1.5 \, GeV^2$ for the second one;
our effective duality interval $\sigma^{CZ}$  for the 
CZ-type DA  appears to be the  average of these two. 
Using the ansatz (\ref{112}) in 
 Eq.(\ref{106}) and integrating over  the 
transverse momentum,  we obtain
\begin{equation} 
F_{\gamma^*\gamma \pi^0}^{LD}(Q^2) = 
\frac{2 \pi}{3} \int_0^1 \frac{\varphi_{\pi}(x)}{x \bar x \sigma} \left [ 1- 
{{1}\over { \left 
( 1+ \bar x \sigma/Q^2 \right )^2 }} \right ] dx .
\label{115} \end{equation} 
This formula  has  correct limits
 both for $Q^2=0$ and large $Q^2$.
 For the asymptotic distribution amplitude, Eq.(\ref{115}) 
produces  the expression
(\ref{98}) coinciding with the
Brodsky-Lepage interpolation formula.
 For the Chernyak-Zhitnitsky DA we get
\begin{equation}  
F_{\gamma^*\gamma \pi^0}^{LD,CZ}(Q^2) = 
\frac{1}{ \pi f_{\pi} } \left \{ 
\frac{1}{1+Q^2/\sigma} - \frac{2 Q^2}{\sigma+Q^2}
 + 12 \, \frac{Q^4}{ \sigma^2} \left [ \left (1+ \frac{2Q^2}{\sigma} \right ) 
\ln \left (1+ \frac{\sigma}{Q^2} \right ) -2 \right] \right \} \, .
\label{116} \end{equation} 
Despite its apparent complexity,
this expression  is very close 
numerically   to the 
simplest interpolation 
\begin{equation}  
F_{\gamma^*\gamma \pi^0}^{int,CZ}(Q^2) = 
\frac{1}{ \pi f_{\pi} (1+  Q^2/\sigma^{CZ})} 
\label{117} \end{equation} 
between the anomaly value at $Q^2=0$ 
and the pQCD result $F_{\gamma^*\gamma \pi^0}^{pQCD, CZ}(Q^2) = \frac53
\, (4 \pi f_\pi /Q^2)$ calculated for the CZ distribution amplitude.

Thus, Eqs. (\ref{98}), (\ref{117}) model the modification
of the basic $I_0$-integral by power corrections.
On the other hand, the modification of $I_0$ 
by radiative corrections is described by Eqs.(\ref{19}),(\ref{21}).
Though we obtained these two types of 
modifications in a completely independent way, it
is tempting to combine them in a single expression. 
A self-consistent, but a rather time-consuming
  way to do this is to calculate the spectral
density $\rho^{quark}(s,Q^2)$ to two loops
and apply the local duality prescription.
Then both the radiative and power corrections would  
result from the same expression.
We leave such a calculation  
for a future  investigation.

In the absence of a completely unified approach,
we can try to get an interpolating formula by  
combining  the two independent 
calculations described above.  
A natural idea is to   write all the one-loop
diagrams in the $b$-representation {\it a l\'a} 
modified factorization 
and  then substitute $\varphi_{\pi}(x,1/b)$
by $\widetilde \Psi (x,b)$ and the Born factor
$\bar \xi K_0(\sqrt{\xi \bar \xi } bQ)$ by the modified version
$\frac14 \xi Q^2 b^2 K_2(\sqrt{\xi } bQ)$.  This will give  a more 
reliable  behavior in the  small-$Q^2$ region where the 
corrections are dominated
by power terms. However, changing the structure of the
Born factor would affect the radiative corrections
and spoil the results at the high-$Q^2$ end,
where one should exactly  reproduce  the pQCD results. 
Since  the perturbative corrections are rather small, 
we   expect that a self-consistent
inclusion of radiative corrections should be 
rather close to a simple    product  
of the nonperturbative $1/(1+Q^2/\sigma)$  factors 
and  perturbative  corrections from 
Eqs.(\ref{19}),(\ref{21}). 
Such a product   gives 
\begin{equation}
F_{\gamma^*\gamma \pi^0}^{as}(Q^2)   \approx  \frac{1}{ \pi f_{\pi}(1+Q^2/s_0)}
 \left\{ 1 -\frac53 {{\alpha_s(Q^2)}\over{\pi}} \right \}
\label{finas} 
\end{equation}
for the asymptotic form of the pion DA, and 
\begin{eqnarray}
&& \lefteqn{F_{\gamma^*\gamma \pi^0}^{CZ}(Q^2)   \approx 
  \frac{1}{ \pi f_{\pi}}
\left \{  \frac1{1+Q^2/s_0} \left\{ 1 -\frac53 {{\alpha_s(Q^2)}\over{\pi}} \right \}
\left [ 1 - \left ({{\ln Q_0^2/\Lambda^2 }
\over{\ln Q^2/\Lambda^2  }} \right )^{50/81}  \right ] \right. \label{fincz} }\\
&& \hspace{4cm} + \left. \frac1{1+\frac35 Q^2/s_0}
 \left\{ 1 -\frac{49}{108} {{\alpha_s(Q^2)}\over{\pi}} \right \} 
\left ({{\ln Q_0^2/\Lambda^2 }
\over{\ln Q^2/\Lambda^2  }} \right )^{50/81}  \right \}\,  \nonumber
\end{eqnarray}
for the case when the pion DA $\varphi_{\pi}(x;\mu)$
coincides with $\varphi_{\pi}^{CZ}(x)$
for $\mu = Q_0 $.  These expressions  have  necessary
 interpolating
properties: in the absence of radiative corrections
they  coincide with the local duality expressions,
while for large $Q^2$, when the power corrections
can be ignored, they  reproduce pQCD  results.
From Fig.4, one can see that 
the  curves for $F_{\gamma^*\gamma \pi^0}^{as}(Q^2)$ and 
$F_{\gamma^*\gamma \pi^0}^{CZ}(Q^2)$ 
(with $Q_0 \approx 0.5 \, GeV$ \cite{czpr}) in this model
are sufficiently separated 
from each other which allows for an unambiguous 
experimental discrimination
between them.

It is instructive to make a more detailed  comparison of 
  the relative size 
of  perturbative $O(\alpha_s)$ 
and  nonperturbative $\sigma/Q^2$  corrections.
Taking $\Lambda  = 200 \, MeV$, we observe that
 the perturbative correction for  the asymptotic 
DA  changes the lowest-order result by $\lesssim 30\%$
for   $ Q^2 \gtrsim  0.5 \, GeV^2$.
This  means that  the pQCD expansion 
for the lowest-twist term in this case is 
self-consistent for $Q^2$ as low as $0.5 \, GeV^2$.
On the other hand,  the power correction
$s_0/Q^2$  exceeds $70 \%$ for all $ Q^2 \lesssim  1 \, GeV^2$.
This clearly indicates  that pQCD results are 
not reliable below $1 \, GeV^2$.  To reduce the ratio
$s_0/Q^2$  to the $20 \%$ level, one should take 
$Q^2 \gtrsim  3 \, GeV^2$. 
This is an illustration of the well-known statement
(see, $e.g.,$ \cite{svz}) that  reliability of simplest 
pQCD formulas 
is limited in first place by power corrections rather 
than by the  increasing value of the QCD running coupling
$\alpha_s(Q^2)$. 
The  crucial  fact here is 
  that the scale $s_0 \approx 0.7 \, GeV^2$ 
determining the deviation from the pQCD  $1/Q^2$
behavior is much larger than $\Lambda^2$. It is also  
much larger than other typical nonperturbative scales like
the square of the
 constituent  quark mass $M_q^2 \sim 0.1 \, GeV^2$
or the average transverse momentum 
$\langle k_{\perp}^2 \rangle$
(in the $LD$-model (\ref{104}), 
$\langle k_{\perp}^2 \rangle^{LD}  = s_0/10 \approx 0.07 \, GeV^2$).
This observation  can be easily explained by the fact
that  $k_{\perp}^2$ present in the modified Born term 
(\ref{106}) is added to $xQ^2$ rather than to $Q^2$.
This  enhances the relative size of  power corrections  by a  
factor like $1/ \langle x \rangle$.
In full accordance 
with the statements made in refs.
\cite{ils,rad}, the onset of the 
$Q^2$-region where the lowest-order
pQCD result is  reliable (in the sense that pQCD gives
a good approximation)  
is determined by the size of the
average virtuality $xQ^2$ of the 
\begin{figure}[t]
\mbox{\vspace{-4cm}
   \epsfxsize=12cm
 \epsfysize=16cm
 \hspace{1cm}
  \epsffile{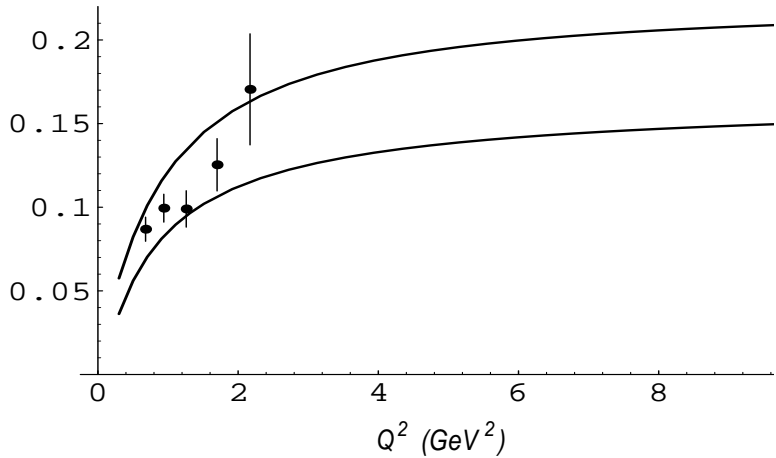}  }
  \vspace{-7.5cm}
{\caption{\label{fig3}
Combination 
$ {\protect \sqrt{2}} Q^2 F_{\gamma^*\gamma \pi^0}(Q^2)/4 \pi$
(measured in $GeV$ and equivalent  to $Q^2 \tilde F_{\gamma^*\gamma \pi^0}(Q^2)$,
with the form factor  $\tilde F_{\gamma^*\gamma \pi^0}(Q^2)$
normalized according to definition adopted in refs.[21,26,38])
as a function of $Q^2$. The lower curve corresponds to 
our model with the 
asymptotic DA (Eq.(5.45)) and the upper one
is based on  Eq.(5.46).  Data  are taken from  CELLO 
collaboration publication [14]. Preliminary CLEO data  
[15] (not shown) are very close to the lower curve.
 }}
\end{figure}
``hard''  quark. 
If  its value is too small, pQCD is unreliable even if
the effective  coupling $\alpha_s$  is negligible 
and perturbation theory 
for the lowest-twist contribution 
is  self-consistent.

\section{Conclusions}

In this paper,  we discussed the 
status of QCD-based  theoretical  predictions 
for the $F_{\gamma^*\gamma  \pi^0}(Q^2)$  form factor. 
 As we repeatedly emphasized,
in this case one deals with a rather favorable situation
when QCD fixes both the $Q^2=0$ value 
(dictated by the axial anomaly) 
and the large-$Q^2$  behavior governed by  perturbative
QCD.  Still, constructing a dynamically supported 
interpolation between 
the two limits, it is very important to 
adequately reproduce at  moderate $Q^2$
the  corrections to the asymptotic pQCD result,
both perturbative and nonperturbative. 

Working  within the framework of the 
standard pQCD factorization approach (SFA),
which allows one to unambigously separate 
the contributions having  different 
power-law behavior at large $Q^2$, 
we gave a detailed analysis of   
the  one-loop coefficient function
for the leading twist-2 contribution.
To explore the role of the transverse 
degrees of freedom, we wrote the relevant Feynman integrals
in the Sudakov representation  
and  showed how the SFA produces  the basic 
building blocks  of  the modified 
factorization approach (MFA) \cite{bottssterman},
such as the Sudakov-type double logarithms $\ln ^2 (b)$
with respect to  the  impact parameter $b_{\perp}$
 which is Fourier-conjugate
to the transverse momentum $k_{\perp}$. 
 The fact that we derived  
the  Sudakov effects  within the lowest-twist 
contribution of the SFA, explicitly demonstrates 
that they should not be  confused with  the higher-twist effects.
In other words, though the Sudakov terms 
are given by integrals over $b_{\perp}$
(or  $k_{\perp}$),  they are purely perturbative
and do not produce  power corrections 
to the lowest-order pQCD result.

Furthermore, we observed that the power corrections $1/Q^2$
due to the intrinsic 
transverse momentum are rather elusive  both within the OPE-type 
factorization and  the light-cone approach of Brodsky and Lepage.
Contrary to naive parton expectations, the simplest
handbag-type diagram 
in both cases does not produce  an  infinite tower  of $(1/Q^2)^n$
terms: such a  series is generated by 
contributions corresponding to physical (transverse)  gluons 
emitted from the hard propagator connecting the photon vertices.
It goes without saying that an explicit summation
of such terms is a formidable task in both of 
these approaches. 
A simpler picture emerges within the QCD sum rule approach
in which the infinite sum over  the 
soft parts of the $\bar q G \ldots G q$ Fock components
is  dual to the $\bar qq$ states generated by the 
local axial 
current.  An important 
observation establishing
the connection between the QCD sum rule
and light-cone approaches is that integrating   the 
invariant mass $s$ of the $\bar qq$-pair over the pion 
duality interval $0 \leq s \leq s_0$  is equivalent
to using the effective two-body wave function $\Psi^{LD}(x, k_{\perp})$.
 The result obtained from the 
local quark-hadron duality ($LD$) ansatz applied to the
lowest-order triangle diagram coincides with the
Brodsky-Lepage interpolation formula \cite{blin},
$i.e.,$ it  reproduces both the $Q^2=0$ value 
specified by the axial anomaly and the high-$Q^2$ 
pQCD behavior with the normalization 
corresponding to the asymptotic distribution amplitude for the pion.
To test the sensitivity to the shape of the
pion distribution amplitude, we proposed 
a model for the effective wave function $\Psi^{LD}(x, k_{\perp})$ which
reduces to the desired DA after the $ k_{\perp}$-integration 
and still  provides  the  correct  limits for the form factor 
both at low and high $Q^2$. 

In our analysis, the  regions  of small  and large transverse
momenta (responsible for power $1/Q^2$ and  
$\alpha_s$ corrections, respectively)
were studied  separately, within the frameworks 
of two different approaches.  In spite of this, the basic 
results written in terms of the $ k_{\perp}$-integrals 
look rather similar.  
A major challenge for a future study  is
the construction of a  unified approach  
in which both the nonperturbative power-suppressed terms
and the perturbative radiative corrections emerge 
from the expansion of the same expression.
  The quark-hadron  duality approach provides
a framework in which such a self-consistent unification 
is guaranteed.  The  {\it only} missing ingredient is 
the perturbative spectral density $\rho^{quark}(s,Q^2)$ 
 at the two-loop level.

There are two further improvements which should
be made in the perturbative part of the problem. 
First, it is necessary to fix 
the argument of the running coupling constant $\alpha_s$.
In our analysis, we either left it unspecified
and estimated the corrections assuming that
$\alpha_s/\pi \approx 0.1$
or took  $\Lambda = 200 \, MeV$
in the 1-loop expression for $\alpha_s (Q^2)$.
However, for a precise  comparison with experimental data, 
estimating  the magnitude  of the $\alpha_s$-correction 
one should explicitly specify the 
$UV$ renormalization scheme,
 fix  the parameter 
$\mu_R$ in the argument of the running coupling $\alpha_s(\mu_R)$
and use the proper   value of the QCD scale 
$\Lambda$. 
A very effective scale-fixing prescription 
is provided by the Brodsky-Lepage-Mackenzie approach \cite{blm}.
To use the BLM prescription, one should calculate 
two-loop pQCD corrections to the coefficient function 
containing quark loop insertions into the gluon propagator.
Another problem is the   inclusion of  the effects due to 
the  two-loop
evolution of the pion distribution amplitude
 \cite{ditradkern,mikhrad,sarmadi}. 
Originally, the relevant  corrections 
expanded in terms of a few  lowest 
eigenfunctions of the one-loop kernel,
were found to be tiny \cite{kmr}.
A recent progress \cite{muller1} in understanding the structure
of the two-loop evolution suggests  that 
higher harmonics cannot be neglected, and the size of the
two-loop evolution 
corrections  is  somewhat larger than estimated in \cite{kmr}. 
However,  our preliminary  numerical estimates 
\cite{2lp} of the effects due to the 
modified evolution developed 
in ref.\cite{muller2}  do not indicate
appreciable changes for  the $I$-integral.

\section{Acknowledgement}

We thank  N. Isgur for encouragement 
and interest in this work; 
R.  Akhoury, V.M. Braun, S.J. Brodsky,  P. Kroll, V.I. Zakharov and
A.R. Zhitnitsky
for stimulating criticism and discussions and V.L. Chernyak
for attracting our  attention to ref.\cite{huang}. 
One of us (A.R.) expresses a special gratitude to G. Sterman
for (in)numerous  discussions and  correspondence 
about  the connection  between 
the standard and modified  factorization approaches.
We thank A.V. Afanasev, V.V. Anisovich, I. Balitsky, W.Broniowski,
W.W. Buck, F. Gross, M.R. Frank,  G.Korchemsky, B.Q. Ma,  L.Mankiewicz,  
M.A. Strikman and A. Szczepaniak for useful discussions.

This work was supported by the 
US Department of Energy  under contract DE-AC05-84ER40150 
and grant DE-FG05-94ER40832 and also by
Polish-U.S. II Joint Maria Sklodowska-Curie
Fund, project number PAA/NSF-94-158.

\end{document}